\documentclass[useAMS,usenatbib]{mnras}

\usepackage{float}
\usepackage{graphicx}
\usepackage{amsmath}
\usepackage{amssymb}
\usepackage{booktabs}
\usepackage{comment}

\def\aap{Astronomy \& Astrophysics}
\def\apj{The Astrophysical Journal}
\def\mnras{Monthly Notices of the Royal Astronomical Society}
\def\apjl{The Astrophysical Journal Letters}
\def\apjs{The Astrophysical Journal Supplements}
\def\pasp{The Astronomical Society of the Pacific}
\def\aj{The Astronomical Journal}
\def\pasj{Publications of the Astronomical Society of Japan}
\def\nat{Nature}
\def\aapr{The Astronomy and Astrophysics Review}
\def\araa{Annual Review of Astronomy and Astrophysics}

\begin{document}

\title[gPhoton: Pulsators]{White Dwarf Variability With gPhoton: Pulsators}

\author[Tucker et al.]{
Michael A. Tucker$^{1,2,\dagger,}$\thanks{Email: tuckerma95@gmail.com},
Scott W. Fleming$^{3,4}$,
Ingrid Pelisoli$^{5}$,
Alejandra Romero$^{5}$,
\newauthor
Keaton J. Bell$^{6}$,
S.\ O.\ Kepler$^{5}$,
Daniel B. Caton$^{2}$,
John Debes$^{3}$,
Michael H. Montgomery$^{7}$,
\newauthor
Susan  E. Thompson$^{3,8,9}$,
Detlev Koester$^{10}$,
Chase Million$^{11}$,
Bernie Shiao$^{3}$
\\
$^{1}$Institute for Astronomy, University of Hawaii at Manoa, 2680 Woodlawn Dr, Honolulu, HI 96822, USA\\
$^{2}$Dept. of Physics and Astronomy, Appalachian State University, 525 Rivers St, Boone, NC, 28608, USA\\
$^{3}$Space Telescope Science Institute, 3700 San Martin D,. Baltimore, MD, 21218 USA\\
$^{4}$CSRA, 3700 San Martin Dr, Baltimore, MD, 21218, USA\\
$^{5}$Instituto de F\'{i}sica, Universidade Federal do Rio Grande do Sul, 91501-900 Porto Alegre, RS, Brazil\\
$^{6}$Max-Planck-Institut f{\"u}r Sonnensystemforschung, Justus-von-Liebig-Weg 3, G{\"o}ttingen\-\,37077, Germany\\
$^{7}$Department of Astronomy, University of Texas at Austin, Austin, TX - 78712, USA\\
$^{8}$NASA Ames Research Center, Moffett Field, CA 94035, USA\\
$^{9}$SETI Institute, 189 Bernardo Avenue Suite 100, Mountain View, CA 94043, USA\\
$^{10}$Institut f{\"u}r Theoretische Physik und Astrophysik, Universit{\"a}t Kiel, 24098 Kiel, Germany\\
$^{11}$Million Concepts LLC, PO Box 119, 141 Mary St, Lemont, PA 16851, USA\\
$^{\dagger}$Participant of the STScI 2015 Summer REU Program
}

\date{Accepted . Received ; in original form }

\pagerange{\pageref{firstpage}--\pageref{lastpage}} \pubyear{2017}

\label{firstpage}
\maketitle
\begin{abstract}
We present results from a search for short time-scale white dwarf variability using \texttt{\texttt{gPhoton}}, a time-tagged database of \textit{GALEX} photon events and associated software package. We conducted a survey of $320$ white dwarf stars in the McCook-Sion catalogue, inspecting each for photometric variability with particular emphasis on variability over time-scales less than $\sim 30$ minutes. From that survey, we present the discovery of a new pulsating white dwarf: WD 2246-069.  A Ca II K line is found in archival ESO spectra and an IR excess is seen in WISE $W1$ and $W2$ bands. Its independent modes are identified in follow-up optical photometry and used to model its interior structure. Additionally, we detect UV pulsations in four previously known pulsating ZZ Ceti-type (DAVs). Included in this group is the simultaneous fitting of the pulsations of WD 1401-147 in optical, near-ultraviolet and far-ultraviolet bands using nearly concurrent Whole Earth Telescope and \textit{GALEX} data, providing observational insight into the wavelength dependence of white dwarf pulsation amplitudes.
\end{abstract}

\begin{keywords}
stars-white dwarfs, stars-oscillations, ultraviolet-stars
\end{keywords}

\section{INTRODUCTION}
\label{introduction}
White dwarf (WD) stars are the final stage in stellar evolution for low and intermediate mass stars; over $95\%$ of stars in the Milky Way will end their lives passively as WDs. These stars are essential to understanding star formation history and offer insight into galactic evolution.  Increasingly, they are also being used to probe the formation and destruction of exoplanets \citep{jura2003,zuckerman2010,koester2014,vanderburg2015}, since their diffusion time scales are much shorter than their evolutionary time scales \citep{paquette1986,alcock1986,koester2009_b}.  A subset of WDs are subject to g-mode instabilities, which manifest in pulsations and enable detailed studies of the WD's internal structure \citep[e.g.,][and references therein]{fontaine08}.

\subsection{Pulsating WDs}

When the upper layers become partially opaque, WDs exhibit non-radial, multi-periodic, \emph{g}-mode pulsations. There are three main types of WD pulsators, each with its own respective spectral type, bulk atmospheric composition, and temperature range: the ZZ Ceti class \citep[DAV, hydrogen-dominated atmosphere,][]{landolt1968}, the V777 Her class \citep[DBV, helium-dominated atmosphere,][]{winget82}, and the GW Vir class \citep[DOV, carbon-oxygen-helium mixed atmosphere,][]{mcgraw79}. Another WD class, the DQ pulsators \citep{dufour07}, whose atmospheres are carbon-dominated with little hydrogen or helium, may represent a fourth class of pulsator \citep{montgomery08,fontaine2008b}, although other effects may be the cause of variability for these objects \citep{montgomery08,dufour2008,williams2013}. 

\begin{figure}
\includegraphics[scale=0.55]{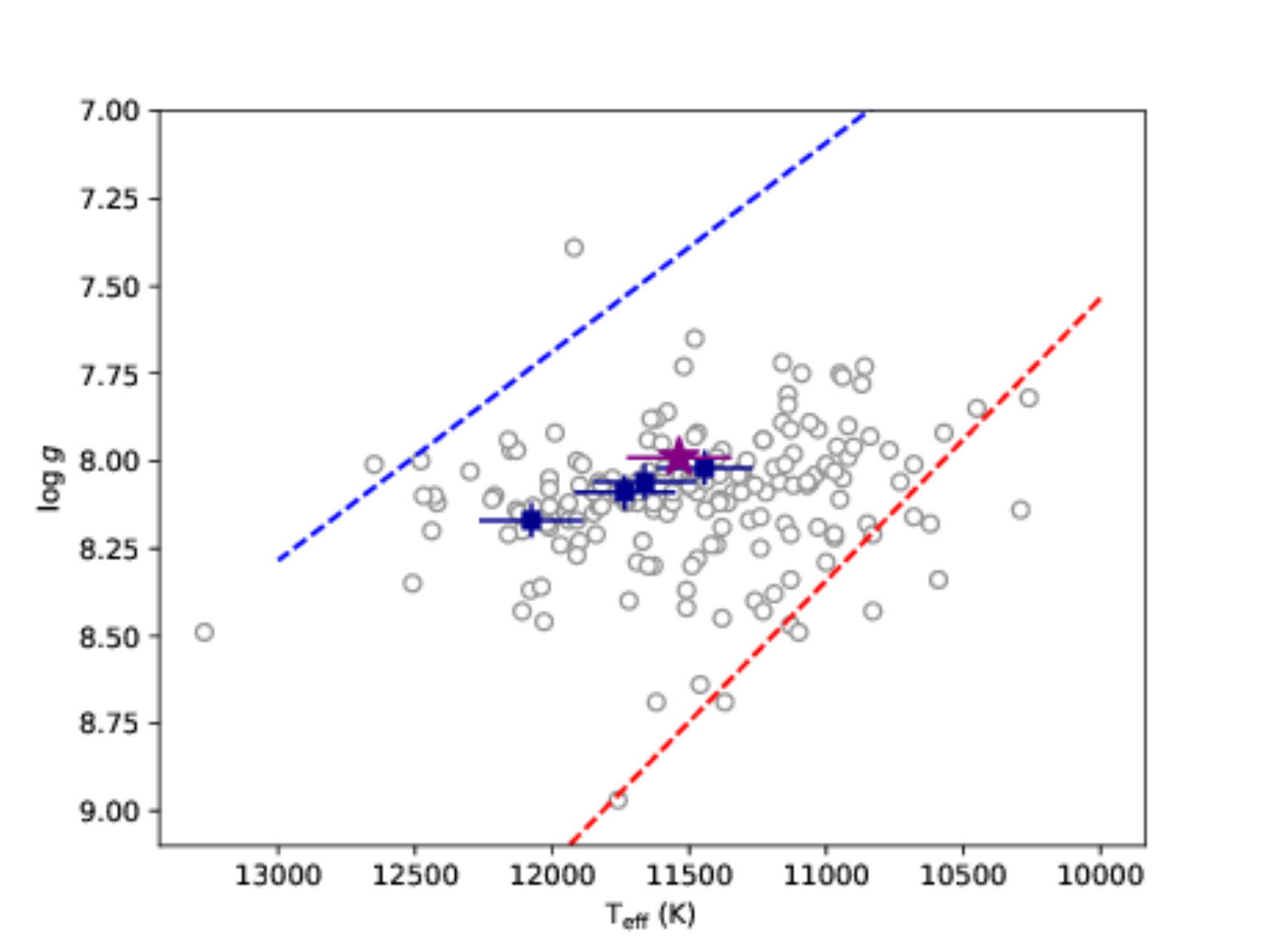}
\caption{Location of ZZ Ceti stars listed in \citet{bognar2016}. Known DAVs are grey circles and DAVs in this paper are plotted as blue squares.  WD 2246-069, the new pulsator described here, is plotted as a purple star. The dotted lines indicate the location of the empirical red and blue edges of the instability strip referred by \citet{tremblay15}.}
\label{insta_strip}
\end{figure}

The DAV class is the most prolific, with over 180 stars confirmed to date,  effective temperatures ranging from $10,500 - 12,500$ K for $\log$ \emph{g} $\sim 8$ \citep{giannis2011}. An up-to-date census of DAVs is shown in Fig. \ref{insta_strip}. The DBV class is driven by the recombination of partially ionized helium, compared to partially ionized hydrogen for DAV stars, and therefore, have a higher effective temperature range of $24,000 - 27,000$ K. These pulsators are significantly less abundant with $23$ DBVs confirmed. GW Vir stars are the hottest pulsator class, having effective temperatures $T_{\rm{eff}} > 75,000$ K and sparsely populated with 22 documented to date. The number of confirmed pulsators in each class is taken from \citet{kepler17}.

Most recently, extremely-low mass DA WDs (so-called ``ELM'' WDs, $\lesssim 0.2 M_{\odot}$) have been found to pulsate due to convective mixing driven by partial ionization in the atmosphere \citep{hermes12,hermes13,bell2017}. \citet{2014PASJ...66...76M} theorize that there is a class of pre-white dwarfs, DAO, that should pulsate between 30,000 - 120,000 K, but these have yet to be confirmed with observation.  More information on WD pulsation characteristics and driving mechanisms can be found in the review by \citet{fontaine08}, and a brief overview of more recent WD pulsation discoveries is provided by \citet{2014IAUS..301..273F}.

\subsection{Peculiar WDs: IR Excess and Metal Lines}

A significant amount of WDs also exhibit metal lines in their atmospheres, classified as "DAZs", with an estimated $\sim 20\%$ \citep{zuckerman2003} of DA WDs falling into this subcategory. These metals are thought to arise from circumstellar accretion of a debris disk, as the majority of DA WDs have gravitational settling time-scales of a few thousand years. These debris disks likely stem from planetesimals that passed within the tidal disruption radius of the WD \citep{jura2003,zuckerman2007,klein2010,debes2012}.

Since these metal lines likely arise from an accretion source (i.e. a debris or gas disk), these WDs often exhibit infrared (IR) flux excess. Studies have detected both gas disks and dust disks encircling WDs, detectable in mid-IR photometry and spectra. Currently $\sim 3-5\%$ of known WDs feature some form of IR excess due to a disk, depending on which survey is used \citep[e.g. ][]{farihi2009,debes2011,barber2014,rocchetto2015,bonsor2017}, although there may be a correlation between WD properties (age, temperature, progenitor) and disk prevalence \citep{koester2014}. \citet{farihi2014} attempted to detect emissions from the population of planetesimals feeding the debris disk, which would likely be at a radius $r \sim 1-100 \; \rm{AU}$, around G29-38 \citep{zuckerman1987} using ALMA and Herschel to no avail.  Most recently, a circumbinary debris disk was found around SDSS J155720.77+091624.6, a binary system composed of a WD and L-type brown dwarf in a 2.27-hour orbit \citep{farihi2017}, adding to the possible configurations of WD debris disks and their planetesimal environments.

Arguably the richest astrophysical WD systems are pulsating WDs that also have metal lines and/or IR excess indicative of circumstellar disks.  These systems allow for precise measurements of the WD interiors, the physics of the accreting material, and the chemical composition of that material.  There are currently four pulsating WDs with documented IR excesses: G29-38 \citep{zuckerman1987}, GD 133 \citep{jura2007}, WD 1150-153 \citep{kilic2007}, and PG 1541+651 \citep{kilic2012}.  Three of these systems also exhibit metal lines in their spectra, the one exception being PG 1541+651, for which there is currently no published optical spectrum. \citet{kilic2012} analysed optical and NIR photometry plus a low-resolution NIR spectrum of PG 1541+651 to derive the properties of the circumstellar debris, including temperature and radius of the disk. 

There have been studies on the effects of pulsations on metal line equivalent widths (EW) with theoretical predictions by \citet{montgomery08} expecting the EW to vary based on amount of accreted material on the surface of the WD ``hot spot''. This can be used to map the distribution of metals, and by assumption the accretion rate, across the surface of the WD. This process was performed by \citet{thompson10} on G29-38, the metallic pulsator prototype, with the conclusion that the accretion may be polar instead of equatorial as previously thought \citep{montgomery08} and raises the possibility that accretion geometry may change with time, although this has yet to be confirmed with future analysis. 

\subsection{WD Pulsation Studies With GALEX}

Since precision asteroseismology requires high cadence photometry over many hours, pulsating WD observations have been predominantly conducted in the optical from the ground. The majority of UV light is blocked by Earth's atmosphere, so most UV data is taken by space-based observatories. In space, the duty cycle is conducive to long time-series photometry, but telescope time itself is at a premium. Although \textit{GALEX}'s photon-counting detectors have the ability to perform time-series photometry, the temporal capabilities were mostly untapped until the release of the \texttt{gPhoton} database \citep{million16}, which facilitated access to calibrated data at the photon-event level.

We present \texttt{gPhoton} light curves in the NUV and, where available, the FUV, for five pulsating WDs whose atmospheric parameters are listed in Table \ref{WDdatatable}. It is worth mentioning the discrepancies between $T_{\rm{eff}}$ and $\log g$ measurements for the same stars across multiple studies. For example, WD 1258+013 was studied most recently by both \citet{giannis2011} and \citet{tremblay11} who derived two different sets of values for $T_{\rm{eff}}$ and $\log g$: $\{11\,990\pm 187 \, \rm K, 8.11\pm 0.05\}$ and $\{11\,400\pm 50\, \rm K, 8.15\pm 0.02\}$, respectively. While the $\log g$ values match within $1\sigma$ uncertainties, the $T_{\rm{eff}}$ do not by a few hundred Kelvin. Additionally, this same star was included in the SSDS DR 7 White Dwarf catalogue \citep{eisenstein13} which modelled the star as having $T_{\rm{eff}} = 11\,099\pm34\,\rm K$ and $\log g = 8.15\pm 0.02$. Some of these disparities can be attributed to the variable nature of the stellar surface for pulsating WDs, especially the fluctuations in temperature. Yet the lack of a clear consensus among three independent studies in two years showcases the need for more comparisons between different sets of spectroscopically derived atmospheric parameters. \citet{Fuchs17} is a recent study of how spectroscopic parameters can change due to various observational factors and how to mitigate such discrepancies. Additionally, comparisons between spectroscopically derived parameters and asteroseismologically derived parameters need to be conducted as evident by the differences for the new pulsating WD 2246-069 (e.g. Table \ref{best-fit}). To maintain consistency, the atmospheric parameters in Table \ref{WDdatatable} were all derived using spectroscopy, taken from the same source when possible and 3D corrections applied corresponding to \citet{tremblay13}. Comparisons between different studies of the same star was facilitated by the Montreal White Dwarf Database \citep[MWDD, ][]{dufour17}.

\begin{table*}
\centering
\caption{Stellar parameters for the 5 DAVs. RA is in hh:mm:ss and DEC is in dd:mm:ss. Effective temperature and surface gravity measurements are 3D corrected per \citet{tremblay13}. Post-correction uncertainties are assumed to be the same as those reported by the reference.}
\begin{tabular}{lcclll}\label{WDdatatable}
WD ID & RA $(\alpha)$ & DEC $(\delta)$ & $\rm{T}_{eff}$ (K) & log(\emph{g}) & Ref. \\\midrule
1258+013 & $13:01:10.52$ & $+01:07:40.05$ &$11\,444\pm176$ &$8.02\pm0.05$ & \citet{giannis2011}\\
1401-147 & $14:03:57.11$ & $-15:01:09.63$ & $12\,077\pm190$ & $8.17\pm0.05$& \citet{giannis2011}\\
1625+125 & $16:28:13.25$ & $+12:24:51.11$ & $11\,662\pm187$ & $8.06\pm0.05$ & \citet{giannis2011}\\
2246-069 & $22:48:40.04$ & $-06:42:45.27$ & $11\,537\pm50$ & $7.99\pm0.02$ & \citet{koester09} \\
2254+126 & $22:56:46.16$ & $+12:52:49.62$ & $11\,735\pm185$ & $8.09\pm0.05$ & \citet{giannis2011}\\\bottomrule
\end{tabular}
\end{table*}

Of the five pulsators in our survey, four are previously known pulsators and one is newly identified. To our knowledge, these are the first UV light curves for the stars in question. Because \textit{GALEX} observations are limited by the orbital parameters of the spacecraft (see Section \ref{datasec}), each continuous light curve is a maximum of $30$ minutes long, although some targets were observed many times over the course of the mission. WD pulsation amplitudes are normally larger in the UV compared to the optical, by up to an order of magnitude, facilitating their detection \citep{bergeron95}. Comparisons of UV-to-optical pulse heights can be used to identify and classify modes when conducting asteroseismology \citep{robinson95,nitta2000,kepler2000,kotak2003,thompson2004}, a notoriously difficult activity even with months of observations in the optical \citep[e.g.,][]{provencal2012}.

In Section \ref{datasec} we describe our target selection and the sources of our UV and optical data. In Section \ref{resultssec} we present the \texttt{gPhoton} light curves of three previously known WDs: WD 1258+013, WD 1625+125, and WD 2254+126. In Section \ref{wd1401sec} we conduct a joint analysis of WD 1401-147 using optical data nearly concurrent with the \texttt{gPhoton} UV observations, comparing the UV-to-optical flux ratio of the strongest frequencies to theoretical models. In Section \ref{wd2246sec}, we present WD 2246-069, a newly identified WD pulsator that also has evidence of an IR excess and metallic pollution via a Ca II feature in its spectrum. Finally, in Section \ref{conclusionsec} we summarize our results.

\section{Data Retrieval and Reduction}
\label{datasec}
\subsection{\textit{GALEX}}
The \textit{Galaxy Evolution Explorer} (\textit{GALEX}) space telescope operated from 2003 until its decommission in 2013, with the primary goal of studying star formation histories in galaxies \citep{morrissey05,martin05}. During its ten-year lifespan, \textit{GALEX} observed $\sim 77\%$ of the sky at various depths in both the near-ultraviolet (NUV, $1771-2831\;\mbox{\AA}$) and far-ultraviolet (FUV, $1344-1786\;\mbox{\AA}$) simultaneously, utilizing two on-board microchannel-plate (MCP) photon-counting detectors. These MCPs record the time and position of each photon event with a time-stamp accuracy of at least five milliseconds. \textit{GALEX} conducted observations on the night side of each orbit, each lasting $1500-1800\;\mathrm{s}$. During each of these `eclipses', \textit{GALEX} could observe one or several parts of the sky, with each location called a `visit.' For deeper surveys, visits lasted nearly the full eclipse (up to 30 minutes), while for shallower surveys each visit lasted only a few minutes, allowing the telescope to point to multiple locations during a single eclipse.  

\textit{GALEX} conducted several survey missions of varying depths throughout its lifespan. The \emph{All-sky Imaging Survey} (AIS) averaged ten $100-200\,\rm{s}$ visits per eclipse with an AB magnitude limit of $\sim 21$. The \emph{Medium Imaging Survey} (MIS) is composed of a single visit taken over an entire eclipse with a magnitude limit of $\sim 23$. The \emph{Deep Imaging Survey} (DIS) coadded visits over approximately $20$ orbits with exposure times totalling $\sim 8\;\rm{h}$. Field selection was constrained by the detector's maximum count rate of $5000\;\rm{counts}\;\rm{s^{-1}}$. \textit{GALEX} was also equipped with low-resolution grism spectroscopy capabilities in both bands.

A comprehensive overview of the original \textit{GALEX} calibration pipeline and available data products is discussed in \citet{morrissey05,morrissey07} and \citet{million16}. In $2011$ NASA ceased direct support for \textit{GALEX} and transferred ownership to the California Institute of Technology for the so-called ``CAUSE'' phase, during which time data was collected and retained by each project's principal investigators until the spacecraft's decommission.

\subsection{\texttt{gPhoton}}\label{gphoton}

\texttt{gPhoton} \citep{million16} is a database and software package that produces high cadence calibrated photometry for \textit{GALEX} by leveraging the 5 ms photon counting nature of the \textit{GALEX} micro-channel plate detectors. The \texttt{gPhoton} database currently contains all direct imaging data from the NASA-funded portion of the \textit{GALEX} mission and is hosted at the Mikulski Archive for Space Telescopes (MAST). A comprehensive overview of the \texttt{gPhoton} data retrieval process, calibration pipeline, and software tools is available in \citet{million16} and on the project webpage\footnote{\url{https://archive.stsci.edu/prepds/gphoton/}}. Our work was conducted using version 1.28.2 of the \texttt{gPhoton} software. We include our \texttt{gPhoton} output files for each target and a script containing our light curve commands as supplemental on-line material.

\subsection{Follow-up Optical Observations}

Follow-up observations of the candidate pulsating WD, DA WD 2246-069 (See \S \ref{wd2246sec}), were taken at Appalachian State University's Dark Sky Observatory (DSO) $32$-inch telescope and the 1.6~m Perkin-Elmer Telescope at Observat\'orio do Pico dos Dias (OPD), in Brazil. At DSO, an Apogee Alta U42 CCD camera with 50~s exposures was used to collect $\sim 8\rm h$ of data across three nights of observations. Due to the faintness of the target and cadence needed to verify variability, the Lum filter was used (a clear filter with an IR blocker to prevent fringing). Images were dark and bias-subtracted and flat-field corrected before performing aperture photometry with Mira Pro X64\footnote{\url{http://www.mirametrics.com}}, Version 8 \citep[Software, ][]{mira1}.

The target was also observed for two nights at OPD, 12 and 15 July 2016, with an Andor iXon CCD and a red-blocking filter (BG40). The CCD has a quantum efficiency of about $60\%$ at $4000$~\AA, a wavelength region where the pulsations have significant amplitude (about tens of mmi, see Eq. \ref{eq:mmi}), so it is suitable for observing pulsating WDs. We observed WD 2246-069 for about 4.25~h on the first night, with an integration time of 15~s plus about 1~s for readout. On the second night we observed for about 5.6~h, with an interruption of 45~min due to clouds. The integration time was again 15~s, but frame transfer was used, reducing the readout to less than 0.1~s.

All the images from OPD were bias-subtracted and flat-field corrected. Aperture photometry was done using the \textit{daophot} package in \textsc{iraf}. A neighbouring non-variable star of similar brightness was used to perform differential photometry. Observation times were corrected to the barycentric celestial reference system to allow the analysis of data for both nights together.

\section{Previously Known Pulsators}
\label{resultssec}

Light curves for each target created with \texttt{gPhoton} show variability with time-scales consistent with asteroseismic pulsations, but due to the short observational baseline ($20-30$ minutes) pulsation frequencies cannot be measured with sufficient precision to identify modes that are present. Brief mode analysis was performed on each target's UV light curves and several signals in the resulting Fourier Transforms match expected WD pulsation periods, and sometimes even match previously reported optical periods at first approximation, however, the short observational baseline prevents any definite conclusions on the validity of these detections. For each target we show the \texttt{gPhoton} NUV and, where available, FUV light curves. We exclude all bins with \texttt{gPhoton} quality flags greater than zero, which indicate one or more warning conditions that, in our experience, are best treated as unusable data. We also exclude any bins that have less than 60\% effective exposure time, i.e., a bin that has less than 6 seconds of data in a 10-second bin.

When analysing variability of targets with \texttt{gPhoton}, it is important to check for false positives due to time dependent sampling of the detector response over the course of a \textit{GALEX} dither, which is under-sampled (and therefore incompletely corrected) by the \textit{GALEX} flat. As such, there can be flux variability strongly correlated to source position on the detector. This can be checked by examining the Fourier spectra of the mean position of events from the centre of the detector (\texttt{detrad}, an output column from \texttt{gPhoton}). This effect is often most severe for bright targets in the non-linear response regime of the MCPs or for objects near the detector edge where the response is less well characterized. For both of these conditions, \texttt{gPhoton} produces warning flags, but such correlations may not be limited to only these situations, and even minor effects matter when looking for low-amplitude variability. We have compared the power spectra of detector position with the flux series and did not find any correlated peaks or aliases.

We note that WD 1645+325 (V777 Her, GD 358), the DBV-class prototype \citep{winget82} has $\{1861,1922\}$ seconds of $\{\rm{FUV}, \rm{NUV}\}$ data in \texttt{gPhoton} and does show pulsation-like variability, but the target is in the non-linearity regime in both bands, where detector effects tend to dominate the light curves.  As such, we omit this target, but note that if a methodology to correct the non-linearity effects is developed these data should be re-examined.

When analysing the light curves for these targets, the fluxes are converted to milli-modulation intensity (mmi) by:
\begin{equation}\label{eq:mmi}
\rm{flux_{mmi} = ((flux / median(flux)) - 1.0) \times 1000}
\end{equation}
\noindent where $1$ mmi $= 0.1\%$ variability.

\subsection{WD 1258+013}
Also known as HE 1258+0123 and V439 Vir, the star is a DA WD first identified as a pulsator candidate on the basis of its location in the HR diagram from spectroscopic effective temperature and surface gravity measurements \citep[$T_{\rm{eff}} = 11\,410$ K, $\log{g} = 8.04$,][]{ber2004}.  Their follow-up optical photometry confirmed its pulsating nature, identifying four peaks in the power spectrum of their 3.3-hour light curve, at periods $P = \{744.6,1092.1,528.5,439.2\}$ seconds, in order of descending amplitude. This target was included in the ensemble analysis of \citet{rom2012}, who included two additional periods in their compilation of known modes ($P = 628.0 \; \rm{and} \; 881.5$ seconds).

\begin{figure}
\includegraphics[scale=0.52]{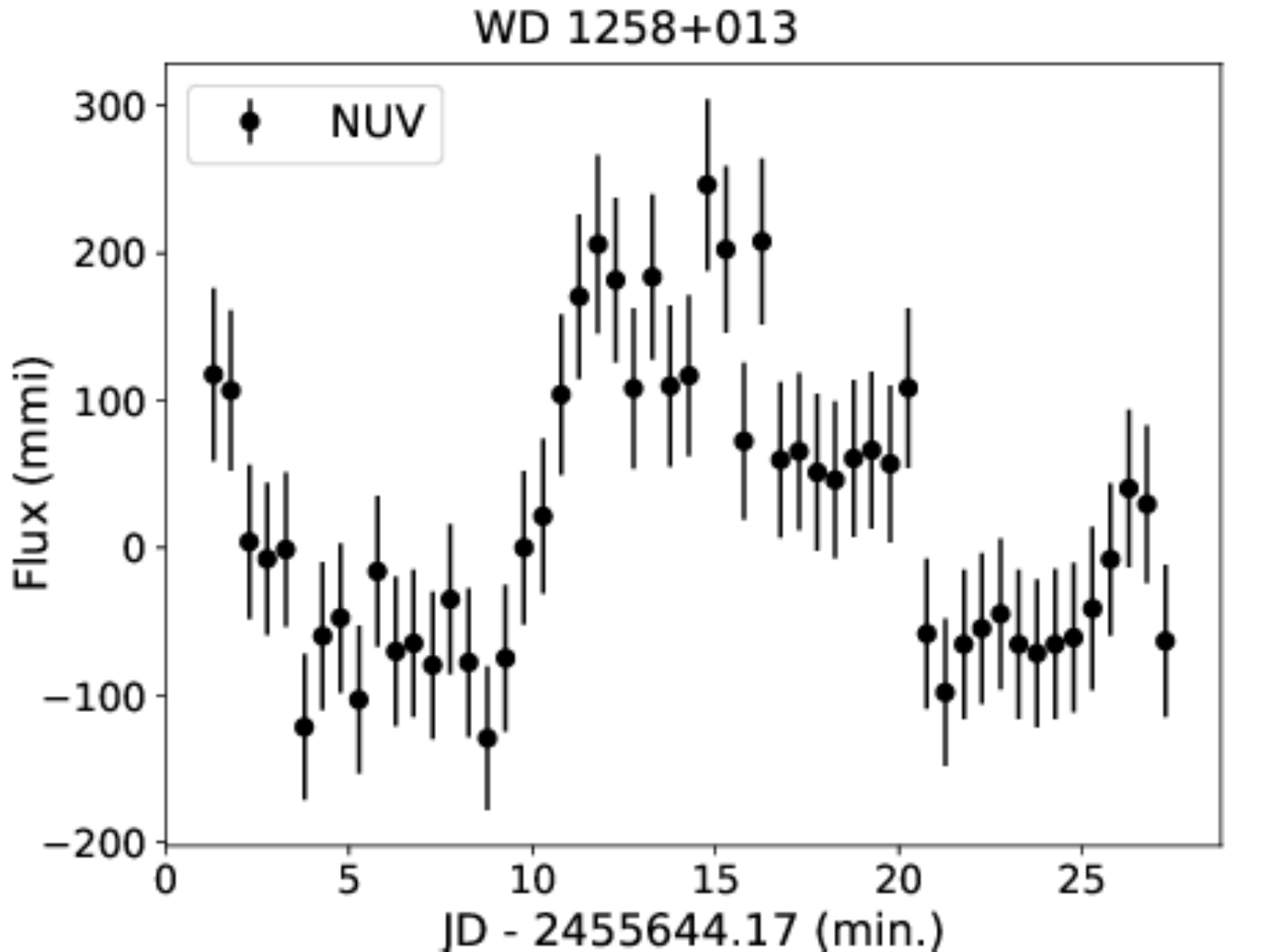}
\caption{\texttt{gPhoton} NUV light curve for WD 1258+013.  Time is in minutes relative to the reference JD.}
\label{1258LC}
\end{figure}

WD 1258+013 is covered by a total of 13 coadd footprints in \textit{GALEX} from 2004-2011, however, only two of these coadds are well-suited for variability searches. Eight of the coadds were part of the shallow AIS survey, each having less than 120 seconds of continuous data. Of the remaining five, three of them are positioned such that the target is too close to the edge of the field-of-view, where edge effects tend to dominate. This leaves two epochs: $\sim 411$ seconds in both FUV and NUV on 17 April 2004 and $\sim 1594$ seconds in NUV on 23 March 2011.  Unfortunately the earlier observations with both bands in 2004 are shorter than any known pulsation periods, so we show only the 2011 data (Fig. \ref{1258LC}). For this target $30\, \rm{s}$ bins were used instead of the $10\, \rm{s}$ bins used for the other targets to maintain adequate counts per bin.

\subsection{WD 1625+125}
Also known as HS 1625+1231, this target was confirmed to be a ZZ Ceti pulsator by \citet{voss06} using photometric selection. Several frequencies were identified, the five largest amplitude periods are \{$862$, $534$, $385$, $2310$, $426$\} seconds in order of decreasing amplitude. This star was also included in the compilation by \citet{rom2012} which featured eight of the ten periods identified by \citet{voss06} as independent modes.

\begin{figure}
\includegraphics[scale=0.52]{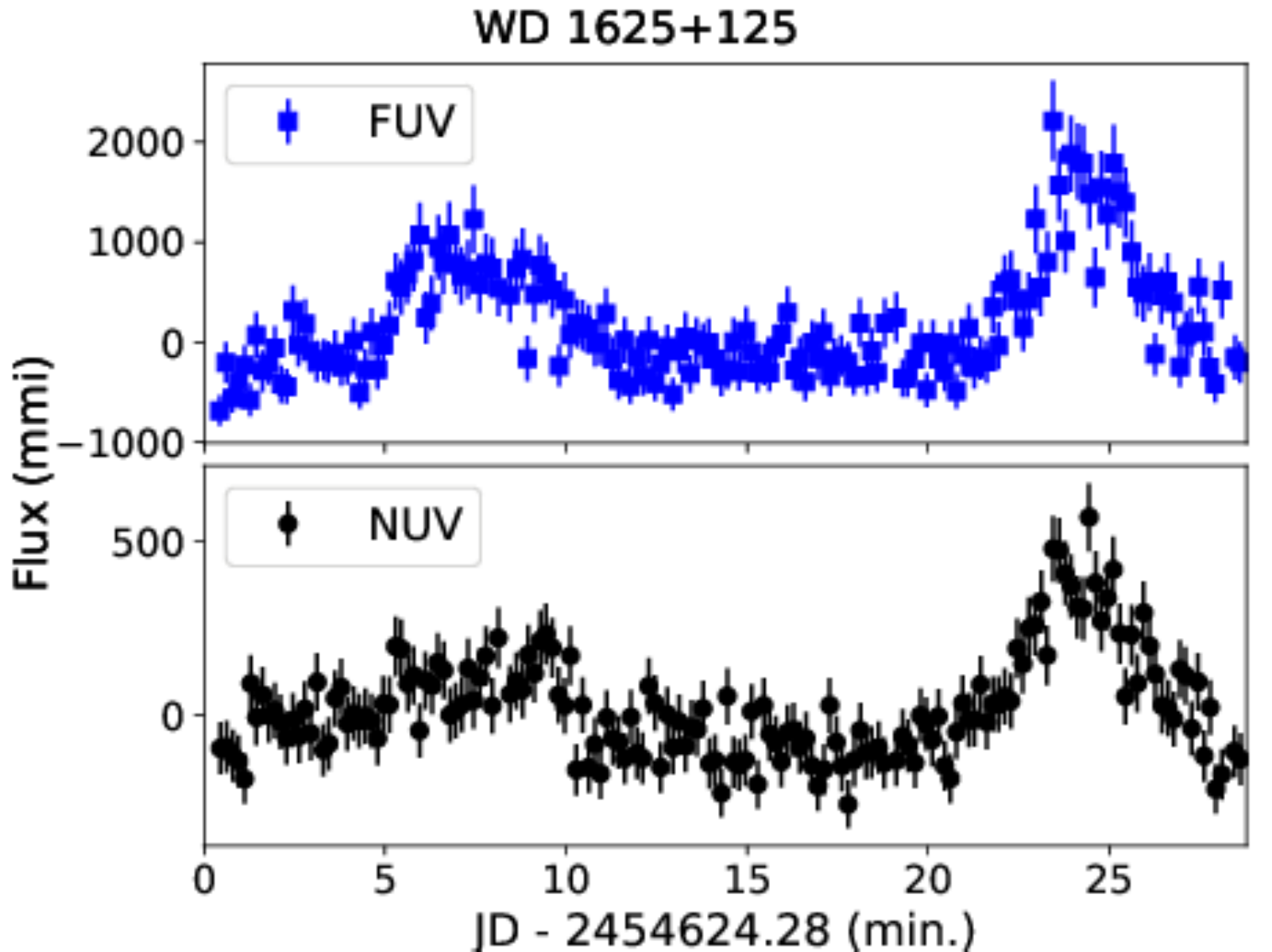}
\caption{\texttt{gPhoton} FUV (top) and NUV (bottom) light curves for WD 1625+125.  Time is in minutes relative to the reference JD.}
\label{1625LC}
\end{figure}

WD 1625+125 is covered by five \textit{GALEX} coadd footprints, four of which contain both NUV and FUV data and one that has only NUV data. Of the five total, all but the last one are too short for asteroseismic purposes. The last, having $\sim 1700\,\rm{s}$ of available data in both bands, was observed on 06 June 2008 and is shown in Fig. \ref{1625LC}. The optical data observations by \citet{voss06} were taken on 15 May 2005, roughly three years before the \textit{GALEX} observations.

\subsection{WD 2254+126}
Also known as GD 244, first speculated to pulsate by \citet{fontaine01} based on optical spectra. Time-series photometry revealed at least $4$ excited modes with periods ranging $200-300\,\rm{s}$.

\begin{figure}
\includegraphics[scale=0.52]{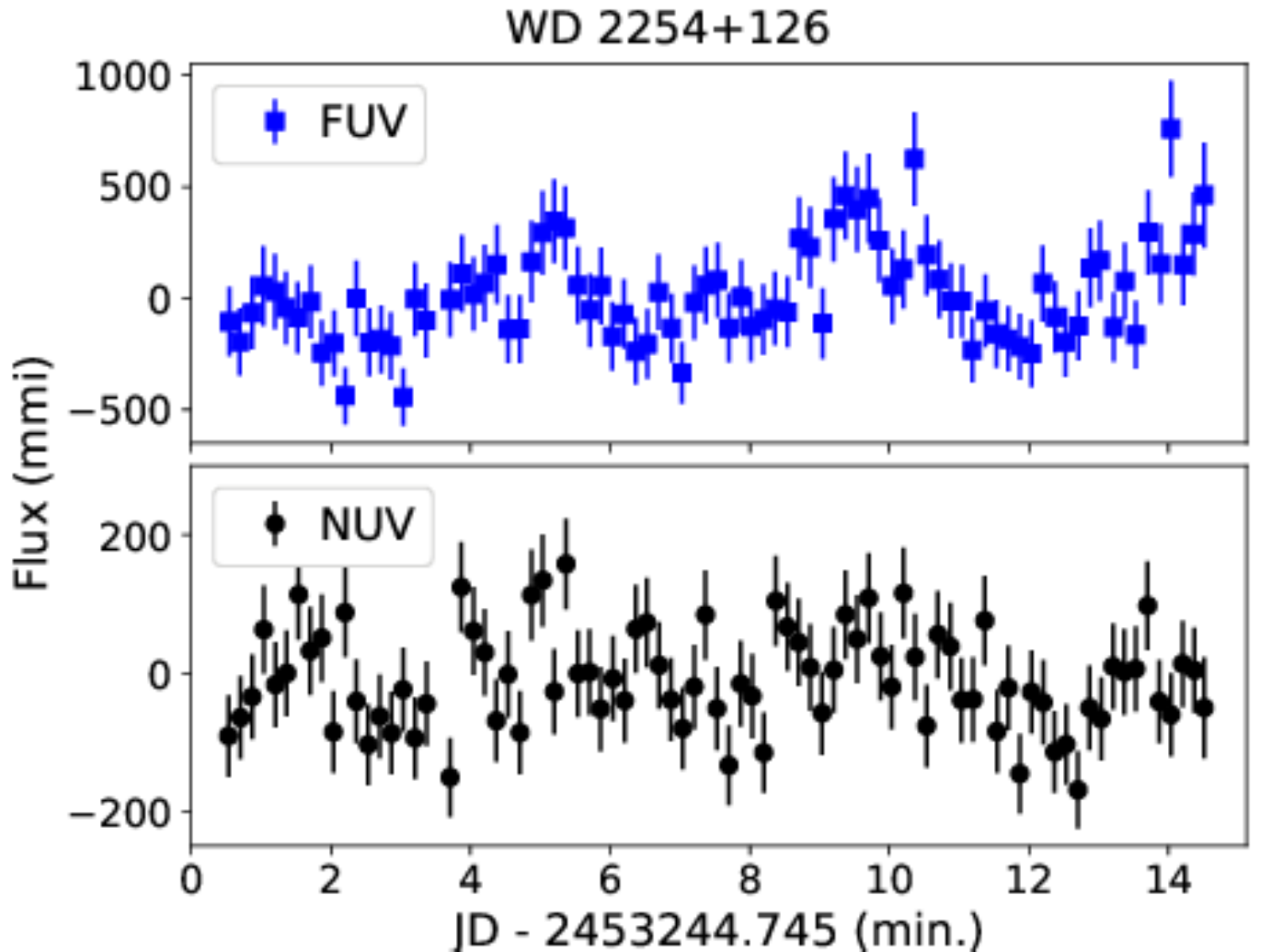}
\caption{\texttt{gPhoton} light curve from Aug. 2004 for WD 2254+126.  The FUV light curve is in blue, the NUV light curve is in black.  Time is in minutes relative to the reference JD.}
\label{2254LC1}
\end{figure}

\begin{figure}
\includegraphics[scale=0.52]{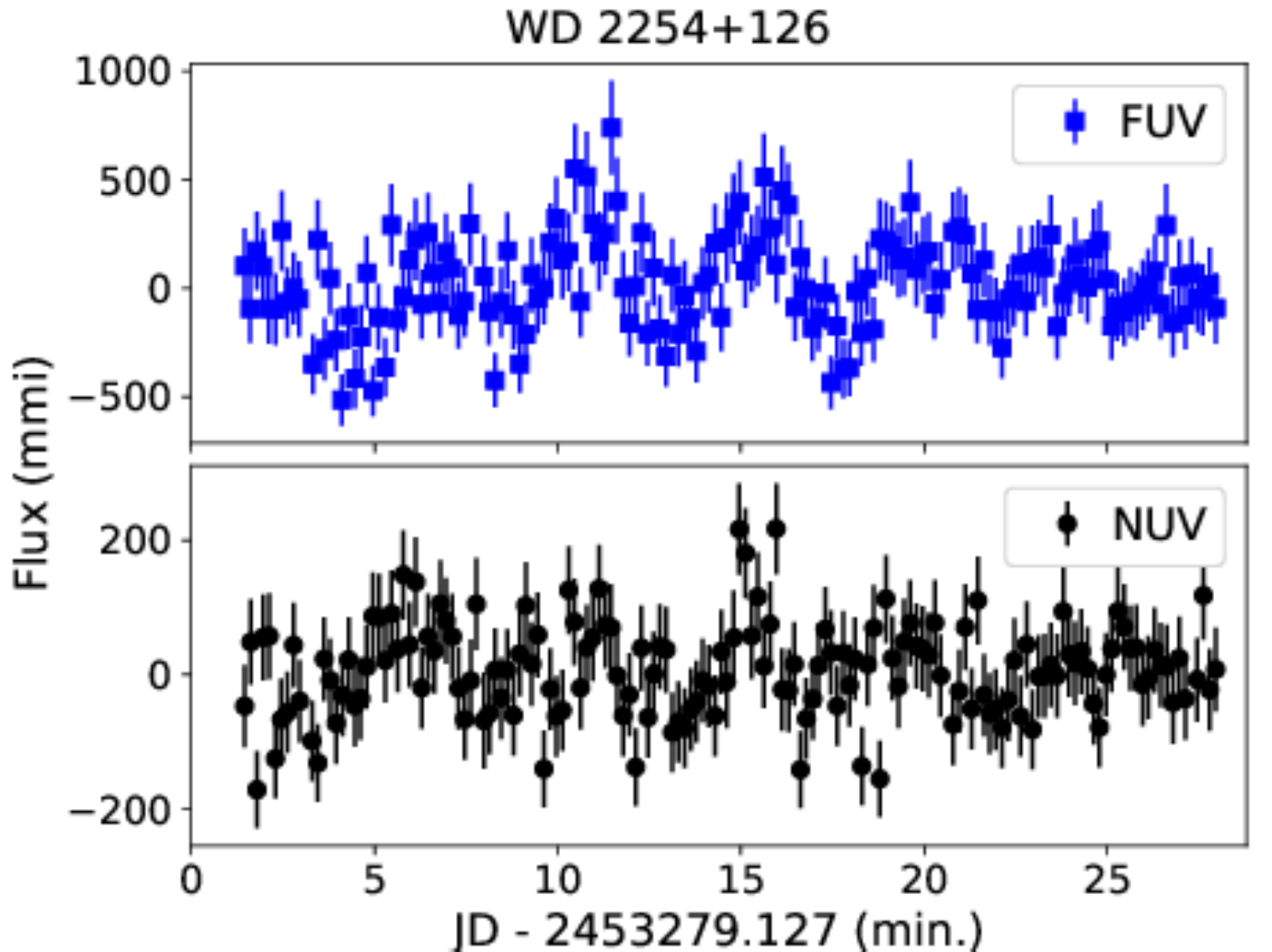}
\caption{\texttt{gPhoton} light curve from Sept. 2004 for WD 2254+126.  The FUV light curve is in blue, the NUV light curve is in black.  Time is in minutes relative to the reference JD.}
\label{2254LC2}
\end{figure}

WD 2254+126 is covered by five \textit{GALEX} coadd footprints, four of which contain both NUV and FUV data and one that has only NUV data. Of the five total, two of them have observation times greater than 500 seconds that can be searched for pulsations: an 845-second observation on 27 August 2004 in both bands and a 1588-second observation on 30 September 2004 in both bands (Figs.\ \ref{2254LC1} and \ref{2254LC2}).

\section{WD 1401-147: A Case of Simultaneous Optical and UV Data from WET and \textit{GALEX}}
\label{wd1401sec}
Also known as IU Vir and EC14012-1446, this star is a well-studied DA WD. It was first identified as a ZZ Ceti from photometric observations in the optical by \citet{stobie95}, who measured five independent modes along with multiple combination frequencies. Further asteroseismic study by \citet{handler08}, using 200 hours of multi-site optical photometry, identified 19 independent frequencies. The most recent observing results come from more than 300 hours of Whole Earth Telescope \citep[WET;][]{WETpaper} observations by \citet{provencal2012}, who also identified 19 independent frequencies, though not all the same ones as those in \citet{handler08}. By combining observations from observatories distributed in longitude, multi-site campaigns like the WET suffer less aliasing than multi-night observations from a single site. \citet{provencal2012} utilize information from the combination modes (amplitudes and phases), mean period spacings and rotational splitting, and non-linear light curve fitting \citep[also utilizing harmonics and combination frequencies;][]{montgomery2010} for mode identification, but these different methods produce some conflicting results. \citet{Chen2014} performed additional asteroseismic analysis based on the combined set of pulsation frequencies of \citet{handler08} and \citet{provencal2012}, putting their own constraints on mode identification.

WD 1401-147 is included in a total of three \textit{GALEX} coadd footprints, two of which contain both FUV and NUV and one contains only NUV data. Two of these are from the shallow AIS survey and have less than 120 seconds of continuous data. The remaining \textit{GALEX} observation of WD 1401-147 is especially interesting, because it was acquired nearly simultaneously with the optical WET data of \citet{provencal2012}, starting only a few hours after the end of the WET observing run enabling a direct comparison of optical to UV pulsation amplitudes.

\subsection{WET Data}
\begin{figure}
\includegraphics[width=1\linewidth]{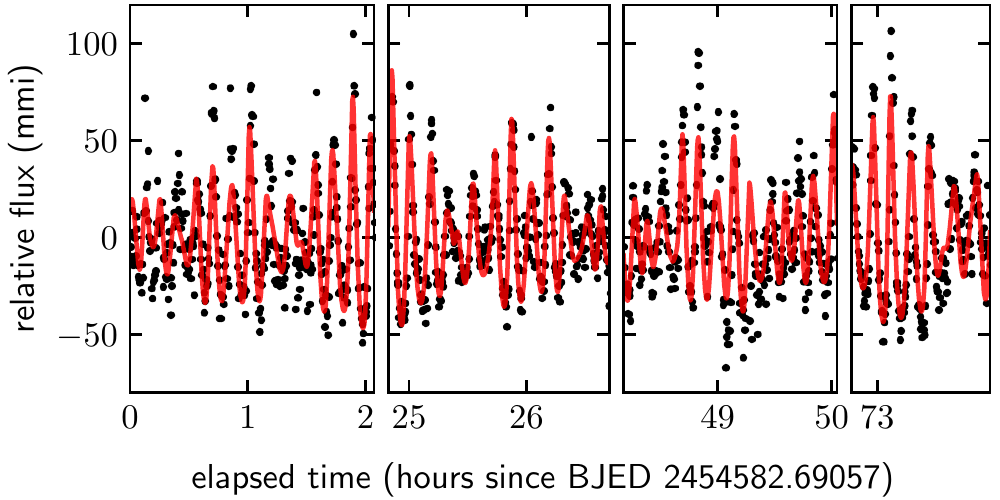}
\caption{Best non-linear light curve fit to the last four nights of WET observations of WD 1401-147 that includes eight independent frequencies (see text). Time on the x axis is displayed in units of hours since the start of WET observations from Brazil at BJED 2454582.69057. Relative flux on the y axis is in units of mmi (Eq. \ref{eq:mmi}).}
\label{WETLC}
\end{figure}

\begin{table*}
\centering
\caption{Fit parameters for WD 1401-147 corresponding to Fig. \ref{1401_LC}. See \S\ref{wd1401sec} for an explanation of the fitting process and mode selection procedure. Provencal ID refers to the modes identified in \citet{provencal2012} for WD 1401-147 and subsequently found in the WET data. }
\begin{tabular}{cccccr}
Freq. $(\mu\rm{Hz})$ & Period (s) & A$_{\rm{BG40}}$ (mmi) & $\ell$ & $m$ & Provencal ID \\
\midrule
$1633.907$ & $612.029$ & $22.2\pm0.6$ & 1 & 1& 1\\
$1774.989$ & $563.383$ & $5.1\pm0.6$ & 1 & 1 & 5 \\
$1887.405$ & $529.827$ & $19.3\pm0.6$ & 1 & 0 & 2 \\
$1891.141$ & $528.781$ & $4.2\pm0.6$ & 1 & -1 & 2b \\
$2504.896$ & $399.218$ & $8.0\pm0.6$ & 1 & 0 & 4 \\
$2508.060$ & $398.714$ & $5.1\pm0.6$ & 1 & -1 & 4a \\
$1548.148$ & $645.932$ & $9.4\pm0.6$ & 1 & 0 & 3 \\
$1521.574$ & $657.213$ & $1.3\pm0.6$ & 1 & 1 & 3a \\
\bottomrule
\end{tabular}
\label{tab:WD1401fit}
\end{table*}

The amplitudes of some pulsation modes of WD 1401-147 were observed to vary on timescales as short as ~days during the 36-day span of WET observations \citep{provencal2012}. To ensure the comparability between data sets, we restrict our comparative analysis to the last four nights of WET data, which were obtained with the Laborat\'orio Nacional de Astrof\'isica (LNA) 1.6-meter telescope at the Pico dos Dias (OPD) Observatory in Brazil through
a red-blocking BG40 filter on 26-29 April, 2008.  The start of the \textit{GALEX} observations begin less than 5 hours after the final exposures of the WET campaign. We utilize the frequency solution of the larger WET dataset to avoid selecting the wrong alias peaks in this subset of the data, and validate the comparability of pulsation amplitudes by demonstrating agreement between model predictions and the GALEX observations

We fit the optical light curve with a script that includes a parametrized treatment of the convection zone's non-linear response to pulsations, which produces harmonic and combination frequencies in the data.  This approach was first introduced by \citet{Montgomery2005}, with subsequent development that includes the application to multi-modal pulsators \citep{montgomery2010}. At each time-stamp, the net effect of pulsations on the emergent stellar spectrum is calculated by integrating over the stellar surface where the local flux is represented by the spectroscopic models of \citet{Koester2010}. We then multiply this net spectrum by an atmospheric transmission model and the BG40 filter throughput to finally produce a simulated optical light curve.

We include eight independent frequencies in our fit to the Brazil WET data, corresponding to IDs 1, 2, 2b, 4, 3, 3a, 4a, and 5 in Table 2 of \citet{provencal2012}, in order of decreasing amplitude. The identified modes are outlined in Table \ref{tab:WD1401fit} of this manuscript. The star was modelled as having an average $T_{\mathrm{eff}}=12{,}077$\,K and $\log{g} = 8.17$ with 3D corrections \citep{tremblay13}. We model the modes based on the $m$ and $\ell$ values supported by the results from non-linear light curve fitting to the entire WET run in the analysis of \citet{provencal2012}. Our best fit to the last four nights of WET observations is displayed in Figure~\ref{WETLC}.

\subsection{\textit{GALEX} Data}

\begin{table}
\centering
\caption{UV-to-optical pulsation amplitude ratios}
\begin{tabular}{lrrrr}\label{ampratios}
spherical degree: & $\ell=1$ & $\ell=2$ & $\ell=3$ & $\ell=4$\\\midrule
$A_{\mathrm{NUV}}/A_{\mathrm{BG40}}$  & 2.26 & 2.68 & 9.67 & 1.14 \\
$A_{\mathrm{FUV}}/A_{\mathrm{BG40}}$  & 6.55 & 8.73 & 47.31 & 1.43 \\
\bottomrule
\end{tabular}
\end{table}

Since the \textit{GALEX} observation was made less than five hours after the final observations of the WET run, we can propagate our best fit to the Brazil data forward to the \textit{GALEX} observations for a unique comparison of UV-to-optical pulsation amplitudes. We run the \textit{GALEX} light curves through the WET analysis pipeline WQED \citep{WQED} to apply identical barycentric and leap-second corrections, so that the optical and UV data can be aligned in phase.

\begin{figure}
\includegraphics[width=1\linewidth]{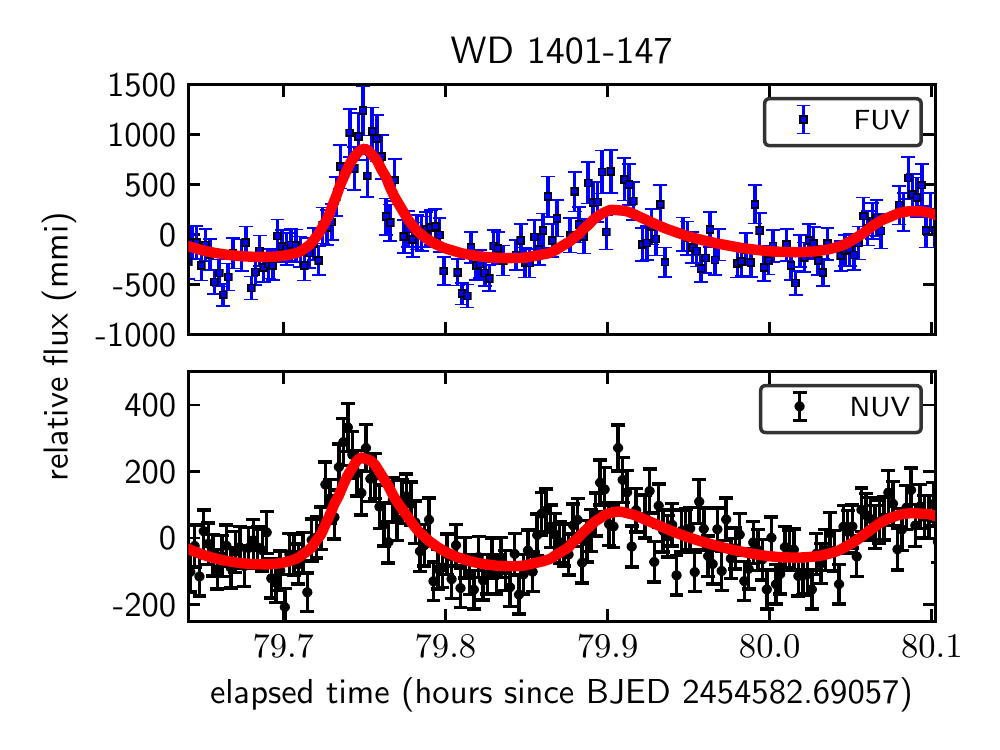}
\caption{\texttt{gPhoton} light curves in FUV (top) and NUV (bottom) for WD 1401-147. The red curves show the predicted UV light curve based on the fits made to the nearly-concurrent optical WET data. Units of mmi are taken from Eq. \ref{eq:mmi}.}
\label{1401_LC}
\end{figure}

Table~\ref{ampratios} gives expected $\ell$-dependent amplitude ratios for independent modes in the \textit{GALEX} passbands relative to the BG40 optical observations. The amplitude ratios are based on pulsating WD models from \citet{Montgomery2005,montgomery2010} and utilizing atmosphere models from \citet{Koester2010}. Restricting the possible $m$ and $\ell$ values to be consistent with the \citet{provencal2012} light curve solution, the results are entirely consistent with the \textit{GALEX} observations. Fig.\ \ref{1401_LC} shows the \texttt{gPhoton} FUV (top) and NUV (bottom) light curves, with the predicted UV light curve based on the best-fit model from the optical WET observations shown as the red curves. This level of agreement supports that the measured pulsations are not $\ell = 3$ or $\ell = 4$, since the UV-to-optical amplitude ratios are very different for these spherical degrees (Table \ref{ampratios}), and reinforces the assumption that WD pulsations are not expected to change on time-scales of a few hours \citep[unless undergoing a newly discovered pulsational outburst phenomenon][]{bell15, bell16, hermes15}.

\section{WD 2246-069, A Newly Identified WD Pulsator}
\label{wd2246sec}
\begin{figure}
\includegraphics[scale=0.52]{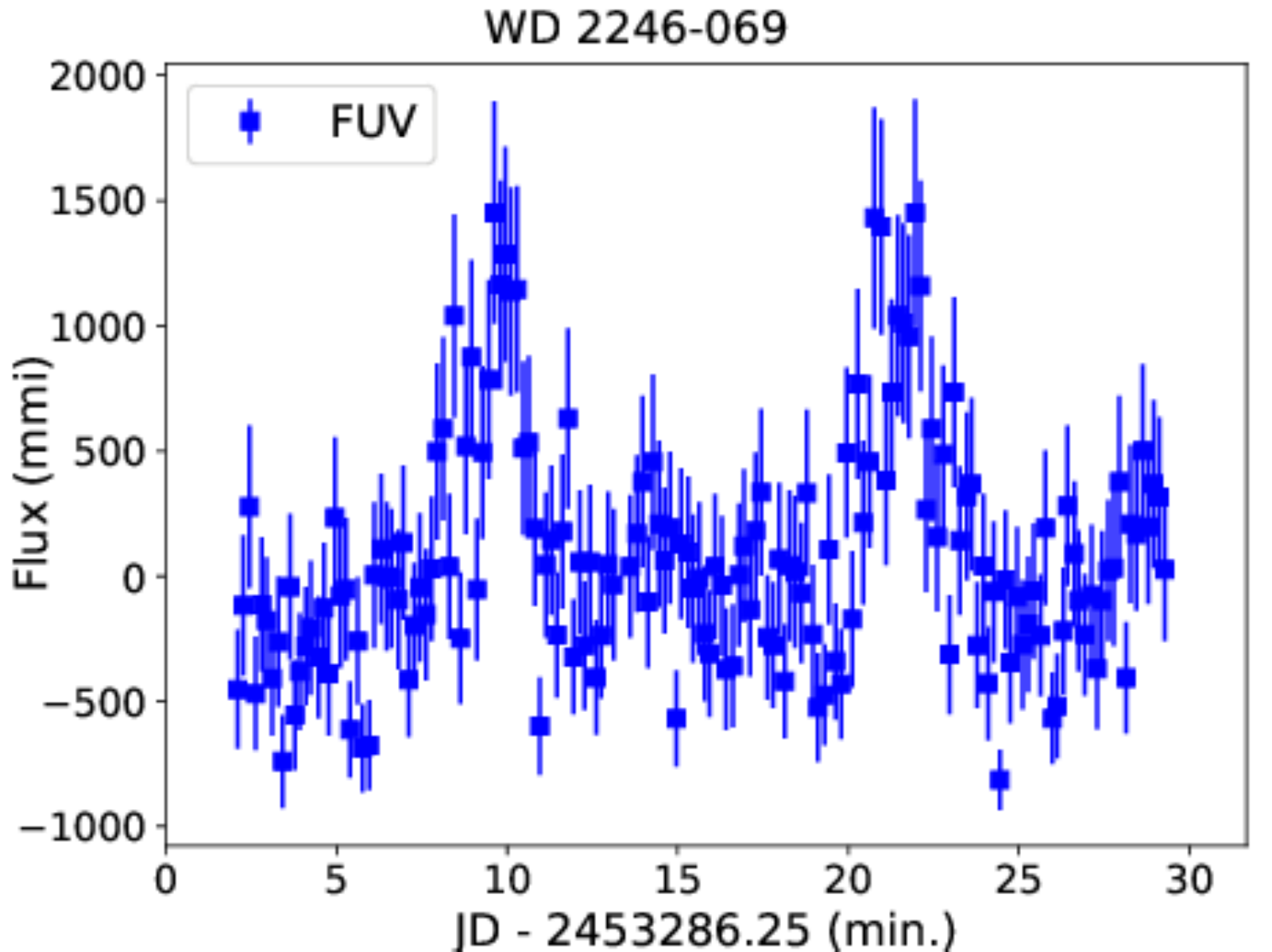}
\caption{\texttt{gPhoton} FUV light curve for WD 2246-069. Time is in minutes relative to the reference JD.}
\label{wd2246_fuv_lcplot}
\end{figure}

\begin{figure}
\includegraphics[scale=0.52]{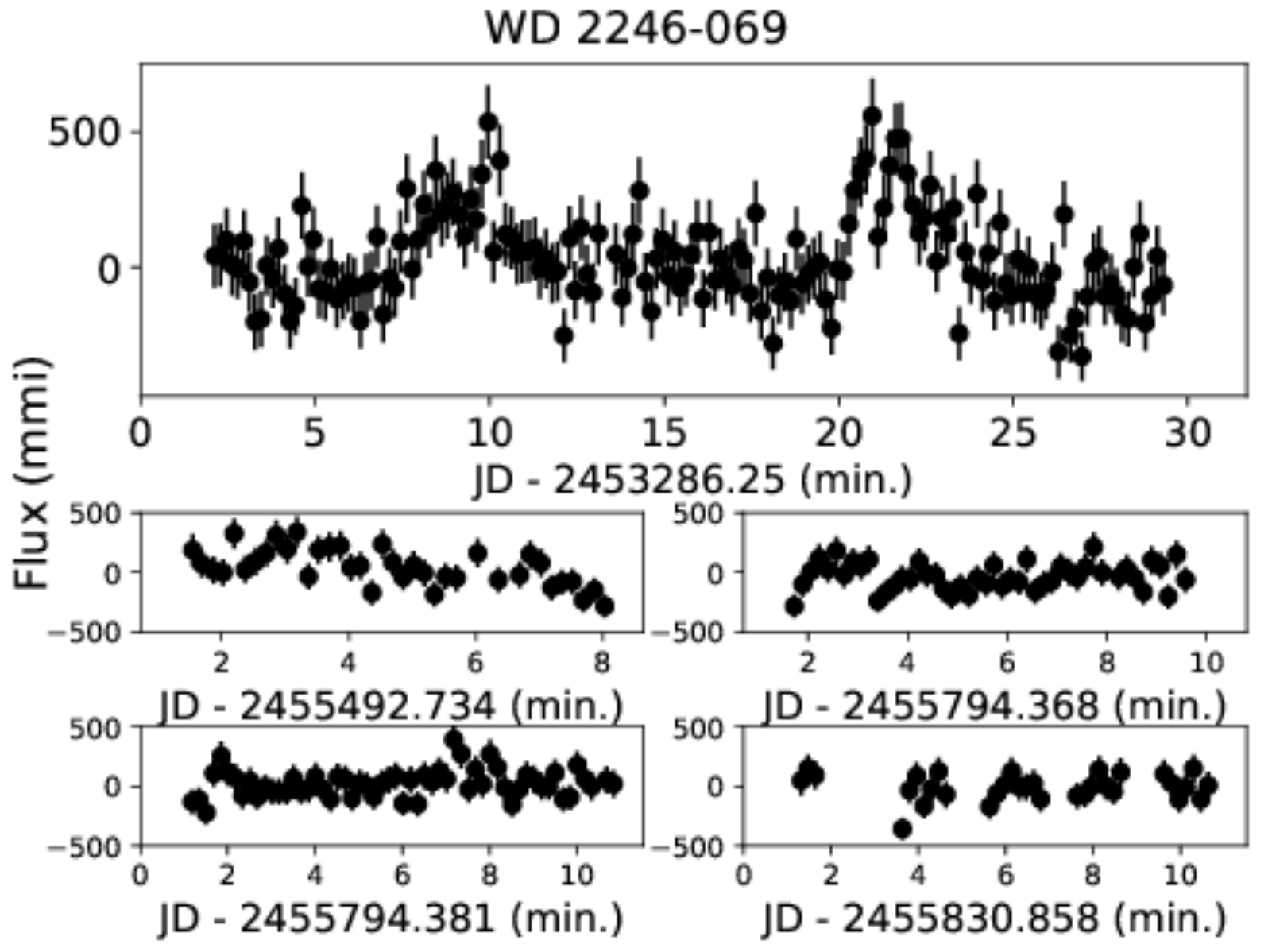}
\caption{\texttt{gPhoton} NUV light curves of WD 2246-069 for all observations with more than 120 seconds of data. Units of mmi are taken from Eq. \ref{eq:mmi}.}
\label{wd2246_nuv_lcplot}
\end{figure}

This survey discovered the pulsations of WD 2246-069 (HE 2246-0658, J224840.063-064244.52), although its atmospheric properties and effective temperature are published in \citet{koester09} and are well within the instability strip (Fig. \ref{insta_strip}).  WD 2246-069 is covered by six \textit{GALEX} footprints, three of which have both FUV and NUV data, while three have only NUV data.  Since this is the first time this star has been identified as a pulsator, we show most of the \texttt{gPhoton} data (Figs. \ref{wd2246_fuv_lcplot} \& \ref{wd2246_nuv_lcplot}).  The one exception is the earliest \textit{GALEX} visit, which was a shallow AIS tile that had only 112 seconds of data. We obtained follow up optical data at DSO and at OPD to confirm the pulsations with longer baselines.

\begin{figure}
\includegraphics[scale=0.64]{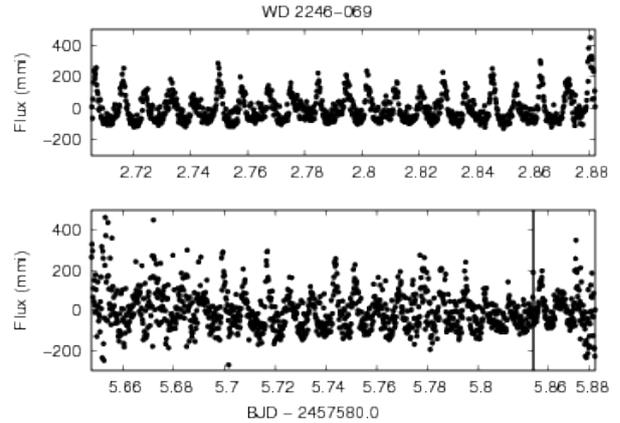}
\caption{Optical light curve obtained at the OPD 1.6~m telescope of WD 2246-069. Top panel shows data for the first night (12 July 2016), bottom panel for the second night (15 July 2016).}
\label{wd2246_opd}
\end{figure}

\subsection{Identification}
Spectra taken by \citet{koester09} put WD 2246-069 at $T_{\rm{eff}}=11\,371 \pm 187 \rm{K}$ and log \emph{g} $= 8.186 \pm 0.06$, right in the heart of the ZZ Ceti instability strip (Fig. \ref{insta_strip}). The \textit{GALEX} UV light curve acquired using \texttt{gPhoton} showed pulsations, but the baseline and gaps between observations make it suitable only for detecting variability.  To verify the pulsating nature of the target, we obtained optical photometry using the 32-inch telescope at the DSO. The DSO light curve exhibits statistically significant variations of $\sim 0.1$ mag. Due to the faintness of the target ($V \approx 16$), instead of the ideal $10\,\rm s$ cadence for pulsating WDs an integration time of $50\,\rm s$ was required, resulting in sampling the light curve at various places between peaks and troughs. However, the DSO data confirms the variable nature of WD 2246-069 and more precise photometry was taken at OPD to conduct asteroseismology and mode identification. The OPD optical light curve is shown in Fig. \ref{wd2246_opd} for the two observed nights. The Fourier transform of the data is shown in Fig. \ref{wd2246_FTopd}. 

\subsection{Asteroseismological analysis}

\begin{figure}
\rotatebox{-90}{\includegraphics[scale=0.33]{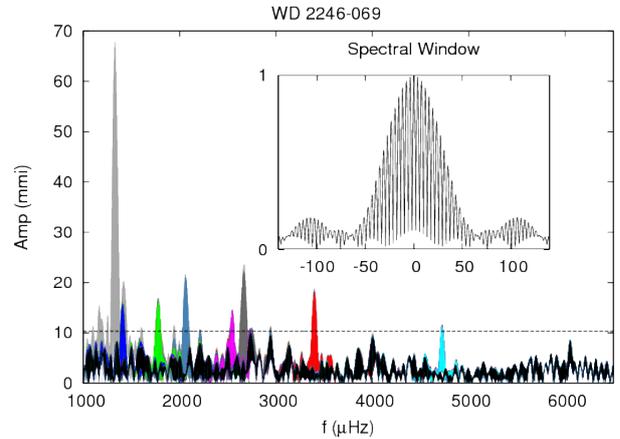}}
\caption{Fourier transform of the WD 2246-069 light curve obtained at OPD (Fig. \ref{wd2246_opd}). The original Fourier transform is shown in light gray; the other colours represent new Fourier transforms after subtracting the found periods. The solid black line shows the last Fourier transform, when nothing is found above the 4$\langle A \rangle$ detection limit (dashed black line).}
\label{wd2246_FTopd}
\end{figure}

Our determination of accurate pulsation frequencies is complicated by extrinsic uncertainties from cycle count ambiguities.  The spectral window displayed in the inlay of Fig. \ref{wd2246_FTopd} depicts the alias structure caused by the three-day gap between OPD observations.  The aliases are finely spaced by roughly $3.9 \, \rm{\mu Hz}$, and the envelope of power drops below $50\%$ relative amplitude near $35\, \rm{\mu Hz}$.  Selecting the intrinsic frequency among these alias signals is non-trivial.  We follow the method of adopting the highest-amplitude alias associated with each pulsational signal.  Photometric noise or signal interference may cause an inaccurate alias of the true pulsation frequency to have the highest amplitude; therefore, the frequencies quoted in Table \ref{list-modes} may be off by some integer multiples of the $3.9\, \rm{\mu Hz}$ extrinsic error; i.e., this is one of many viable frequency solutions.  We quote only the formal, intrinsic uncertainties on each signal frequency in Table \ref{list-modes} \citep{montgomery1999}.

From the Fourier analysis applied to the optical data from OPD we obtain 12 signals above $4\langle \rm A \rangle$, where $ \langle \rm A \rangle$ is the average amplitude of the Fourier transform. Five signals are identified as independent modes, while the remaining seven signals are identified as harmonics or combination frequencies that arise from non-linearities in the light curve.  We used the Period04 Fourier analysis software \citep{lenz05} to derive our frequency solution.  With each mode selected, we improved our best-fit, least-squares solution and computed a new Fourier transform of the residuals to search for additional significant signals.  We included significant harmonic and combination frequencies as soon as their parent modes were selected, otherwise we selected the pulsation signals in decreasing order of amplitude.  We selected the highest peak in the alias structure of each independent pulsation mode for the solution listed in Table \ref{list-modes}, while forcing the harmonic and combination frequencies to be exact multiples and sums of the pulsation frequencies.  For the dominant signal near $1330.3\, \rm{\mu Hz}$, the two highest aliases had essentially the same amplitudes.  While the lower-frequency alias at $1326.4\, \rm{\mu Hz}$ was marginally stronger, we selected the $1330.3\, \rm{\mu Hz}$ peak because it produced harmonic and combination frequencies that match closer to the observed highest-amplitude aliases.

\begin{table}
\centering
\caption{Frequency spectrum from optical observations at Pico dos Dias Observatory corresponding to Fig. \ref{wd2246_FTopd}.}
\begin{tabular}{llcrr}\label{list-modes}
    & Freq.[$\mu$Hz]  & $\Pi_{Obs}$ [s]&  Amp [mmi] & Note \\\midrule 

f1  & $1330.34\pm0.06$ & $751.69 \pm 0.04$ & $69.19 \pm 1.92$ &       \\
f2   & $2060.45 \pm 0.19$  & $485.33 \pm 0.05$   & $20.50\pm 1.92$      & \\
f3  & $1778.07\pm 0.24$ & $562.41\pm 0.08$ & $16.20\pm 1.92$ & \\
f4  & $1406.62\pm 0.25$ & $710.92\pm 0.13$ & $15.87\pm 1.92$ & \\
f5   & $2538.55\pm 0.26$ & $393.93\pm 0.04$ & $14.90\pm 1.92$ & \\
\hline
f6   &  $2660.67\pm 0.16$ & $375.85\pm 0.03$   & $23.49\pm1.92$    &  2f1 \\
f7   & $3390.78 \pm 0.22$ & $294.92\pm 0.02$ & $17.62\pm 1.92$ & f1+f2 \\
f8   & $4721.12\pm0.35 $ & $211.81\pm 0.02$ & $11.09\pm 1.92$ & 2f1+f2 \\
f9   & $2736.96\pm 0.40$ & $365.37\pm 0.06$ & $9.80\pm 1.92$ & f1+f4\\
f10   &  $3991.01 \pm 0.46$    & $250.56 \pm 0.03$       & $8.49\pm 1.92$    &  3f1 \\
f11   & $3108.40\pm 0.71$ & $321.71\pm 0.08$ & $5.36\pm 1.92$ & f1+f3 \\
f12   & $6051.46\pm 0.46$ & $165.25\pm 0.02$ & $8.41\pm 1.92$ & 3f1+f2 \\

\bottomrule
\end{tabular}
\end{table}
Using the modes classified as independent modes in Table \ref{list-modes} (f1 to f5) we perform an asteroseismological fit. We employed an updated grid of ZZ Ceti models from \citet{rom2012, rom2013} and minimize the quality function $S$ defined as:

\begin{equation}
S =\sqrt{ \sum^{N}_{i=1} \frac{[\Pi_k^{th} -\Pi_i^{obs}]^2 \times w_i }{\sum^N_{i=1}wi}}
\end{equation}

\noindent where $N$ is the number of observed modes and $w_i$ are the amplitudes. Since we do not have an identification of the harmonic degree from observations, we penalize the modes with $\ell =2$ by a weight $N_{\ell=2}/N_{\ell=1}$, where $N_{\ell=2}$ and $N_{\ell=1}$ are the number of quadrupole and dipole modes in a given model, respectively. As a result of our seismological study, we obtain a best fit model characterized by a stellar mass of $0.745\pm 0.020 \; M_{\odot}$ and an effective temperature of $T_{\rm eff}=11\,649\pm160\,\rm K$. The structure parameters characterizing our best fit model are listed in Table \ref{best-fit}.  For comparison, we include the spectroscopic parameters and stellar mass determined from the mass-radius relations in \citet{rom2012, rom2013} for C/O core WDs. Also, we list the theoretical periods with the corresponding harmonic degree $\ell$ and radial order $k$.

The seismological effective temperature is somewhat higher, but in good agreement with the spectroscopic determinations. On the other hand, seismology gives a stellar mass $\sim 25 \%$ higher than the spectroscopic value. Note that, as discussed in \S\ref{introduction}, the atmosphere values can change with time due to the variable nature of
the stellar surface for pulsating WDs. Also considering that the available spectra have low S/N and their quoted uncertainties are the internal uncertainties from the fitting procedure, we suspect that the real uncertainties in the spectroscopic determination are underestimated. 

Conversely, due to the aforementioned $3.9 \; \mu \rm{Hz}$ extrinsic uncertainty in the determination of independent modes, the asteroseismological fit is not considered a perfect solution either. While great care was taken to select the correct peaks in the FT (Fig. \ref{wd2246_FTopd}), selecting the right peaks among the aliases is a non-trivial task. Taking the periods determined in Table \ref{list-modes} at face value, the quality function value for the representative model is $S=0.999\, \rm s$, with a mean difference of 0.85 s between theoretical and observed periods. We consider this model a good fit to the data, but provide both sets of parameters in Table \ref{best-fit} due to the uncertainties in both the spectroscopic and asteroseismic fits.

\begin{table}
\centering
\caption{Structure parameters characterizing the asteroseismological best fit model for WD 2246-069. Also listed are the theoretical periods corresponding to the best fit model, along with the harmonic degree and radial order.}
\begin{tabular}{lll}\label{best-fit}

Param. & Spect. & Seism.\\\midrule
  
$M/M_{\odot}$ & $0.599\pm 0.015$ & $0.745 \pm 0.020$   \\
$T_{\rm eff}$ & $11\, 537\pm 50 \, \rm K$ & $11\, 649 \pm 160 \,\rm K$      \\
$\log g$ & $7.99\pm 0.02$ & $8.251 \pm 0.028$  \\
$M_{\rm H}/M_{\odot}$ & $\cdots$ & $4.26\times 10^{-6}$  \\
$M_{\rm He}/M_{\odot}$ & $\cdots$ & $6.61 \times 10^{-3}$  \\
$S$ & $\cdots$ & $0.999\, \rm s$ \\       
\hline\hline
$\Pi_{Theo}$ &  $\ell$  & $k$ \\
\hline
751.7758 & 1 & 16 \\
561.5157 & 2 & 21 \\
392.2792 & 2 & 14 \\
712.9233 & 1 & 15 \\
484.4741 & 2 & 18 \\
\bottomrule
\end{tabular}
\end{table}

\subsection{Spectral Feature}
\label{sec:spectrum}

\begin{figure}
\includegraphics[width=1.05\linewidth]{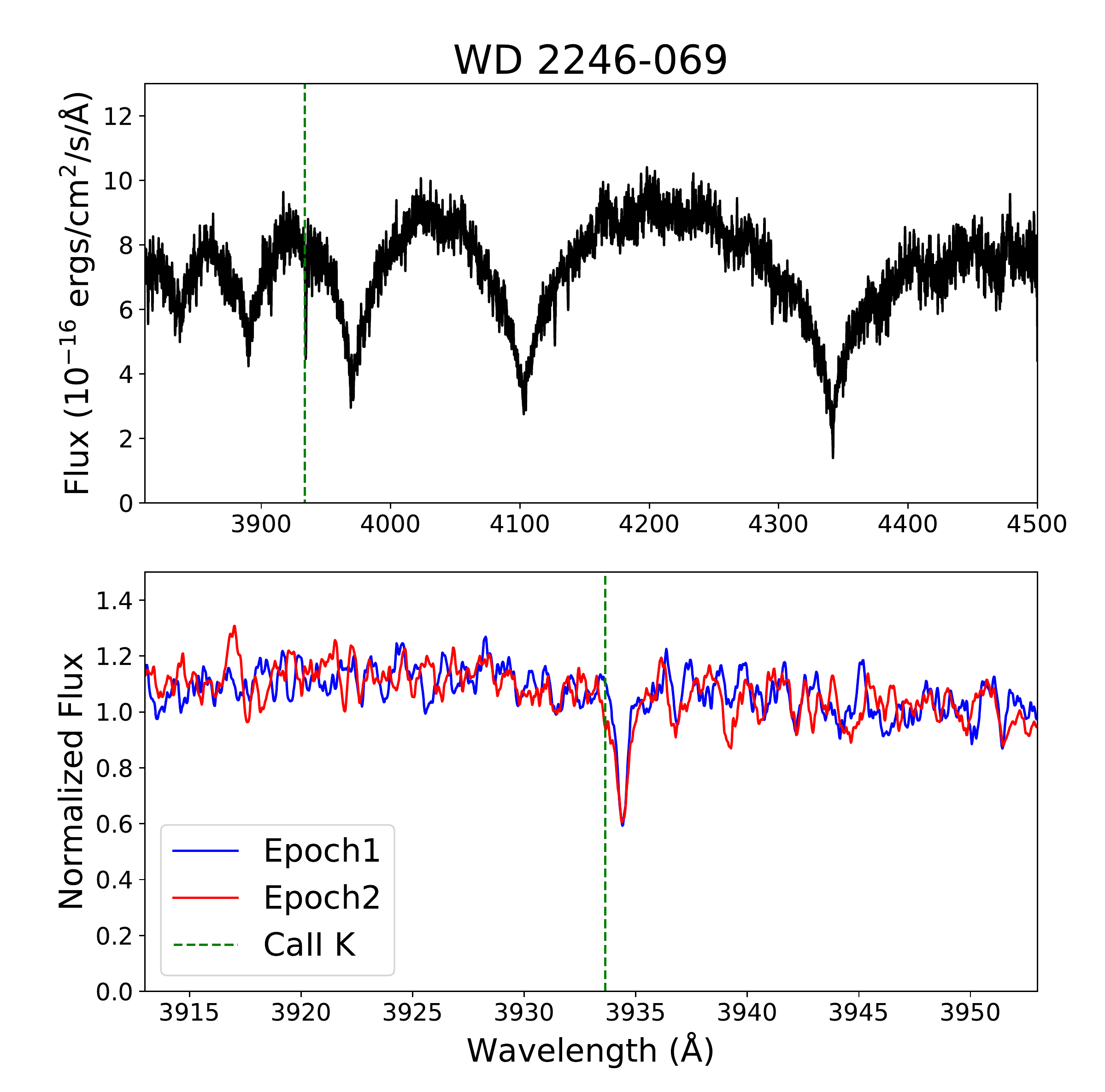}
\caption{\textit{Top}: One of the UVES spectra of WD 2246-069 showing broad H absorption lines typical of a DA WD. A boxcar smoothing was applied, and a few of the worst residual artefacts linearly interpolated over, for display purposes. The green, dashed vertical line in both panels indicates the rest position of the Ca II K line. \newline 
\textit{Bottom}: Both UVES spectra centred on the Ca II K absorption line. They have been heliocentric (but not gravitational redshift) corrected, and show no major radial velocity shift between the two epochs.}
\label{2246spectra}
\end{figure}

\begin{figure}
\includegraphics[width=\linewidth]{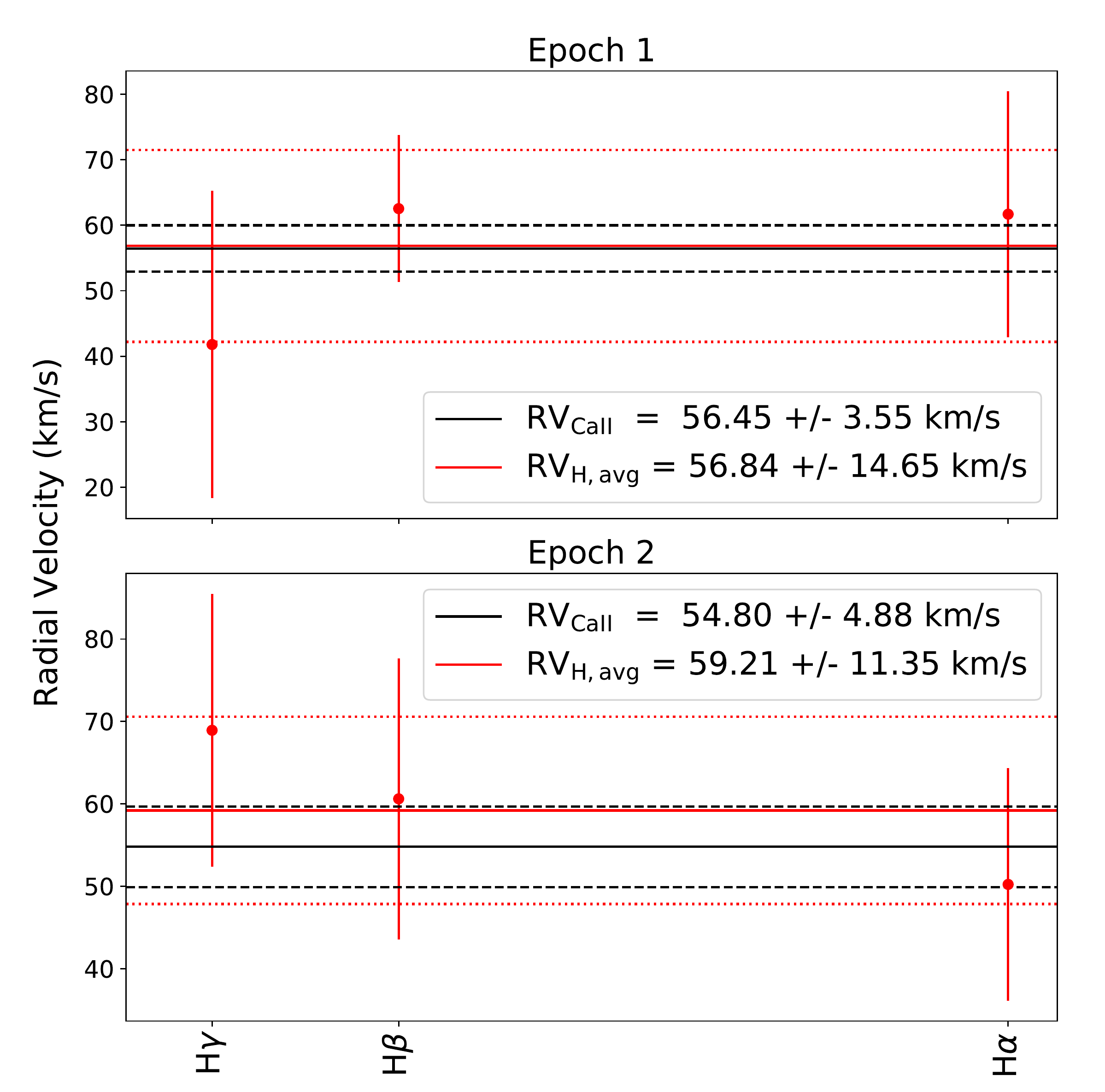}
\caption{Comparisons between the RV of the Ca II absorption feature at $3933.6\rm$\AA  with the RV of the Balmer lines. Red points are RVs from H$\alpha$, H$\beta$ and H$\gamma$. The red solid line and red dotted lines indicate the average Balmer line RV and $1\sigma$ uncertainties, respectively, for each epoch. The black solid line and black dotted lines indicate the RV of the Ca II line and $1\sigma$ uncertainties, respectively. The RV of the CaII feature and the H lines match within $1\sigma$ indicating, among other evidence, that the Ca II line is photospheric and not an ISM feature. The spectra are not corrected for gravitational redshift due to the mass uncertainty.}
\label{fig:RVshifts}
\end{figure}

We obtained the spectra observed as part of the SPY Type Ia progenitor search \citep{SPYpaper} and later re-analysed by \citet{koester09} to determine the atmospheric properties of the observed WDs. These spectra were taken using UVES \citep{dekker2000} on the VLT and retrieved from the ESO data archives to check for radial velocity variability or the presence of any potential metal lines.  We find a Ca II K absorption line feature in both spectra, taken roughly a year apart (Fig.\ \ref{2246spectra}).  There is no evidence of any large RV shift between the two epochs, but given the long baseline and the fact that there are only two spectra we can not definitively rule out RV variability, which could indicate the presence of a bound, unseen companion.

After applying heliocentric corrections (as recorded in the ESO FITS headers) we checked to see whether the RV shift of the Ca II K line matches the cores of the Balmer lines.  We fit a Gaussian to both the Ca line and the centres of the Balmer lines ($\sim \pm7 \textrm{\AA}$ around the rest wavelength).  Fig.\ \ref{fig:RVshifts} shows the RV shift of H$\alpha$, H$\beta$ and H$\gamma$ (red points) compared to the Ca II line for both epochs.  In this figure, the average across all three Balmer lines is also shown as a horizontal red line, and the associated 1$\sigma$ uncertainties are shown as the dashed red lines.  The Ca II RV shift is the solid black line, with the 1$\sigma$ uncertainty shown as the dashed black line.

We find that the Balmer line shifts match the Ca II line to within the fit uncertainties, which are on the order of 10-15 $\rm{km \; s^{-1}}$ due to both the intrinsic width of the Balmer lines and the low signal-to-noise of the spectra (median signal-to-noise of $\sim 5$ across the spectral orders).  We also checked the RV shifts of the other Balmer lines available in the ESO spectra and find they are also consistent with the Ca II RV shift to within the fit uncertainties, however, the fit uncertainties are larger and don't impact the average result across all the Balmer lines significantly, so they are not shown in Fig.\ \ref{fig:RVshifts} for clarity.

A key question is whether this Ca II feature is coming from the WD or is an ISM feature.  There are only a handful of pulsating WDs with metal lines present in their atmospheres \citep[e.g., G29-38, GD 133, WD 1150-153;][]{koester1997, koester05}.  There are a couple lines of evidence that the Ca II feature seen in the ESO spectra are coming from WD 2246 and is not an interstellar medium (ISM) absorption line.  First and foremost, the Balmer line shifts agree with the Ca II RV shift (Fig.\ \ref{fig:RVshifts}).  Second, we checked the Local ISM Kinematic Database\footnote{\url{http://lism.wesleyan.edu/LISMdynamics.html}} \citep{redfield08} and find only one LISM cloud (the LIC) that intersects our target's coordinate, and that has a velocity of $-3.2 \; \pm \; 1.4 \; \rm{km \; s^{-1}}$, which is $\sim3 \sigma$ from our measured Ca II RV even when the higher gravitational redshift of $-44~\rm{km}\,\rm s ^{-1}$ (corresponding to $M_{WD} = 0.745M_\odot$) is applied.  Given the proximity of the WD ($\sim 50$ pc based on our SED fit in Section \ref{irexcess}), only a local ISM cloud could be responsible for such a feature if it did not originate from the WD itself.

\citet{koester09-2} determined the diffusion time-scale for the metal polluted pulsator G29-38 to be $\sim 0.8\,\rm{yr}$. G29-38 is very similar to WD 2246-069 with a slightly lower $\log \,g$ so we take this diffusion time-scale as a upper constraint for WD 2246-069. The presence of metals on the surface often indicate an accretion source due to the short lived nature of such features.  All existing evidence suggests this Ca II feature is coming from the WD itself.  Additional high signal-to-noise, high-resolution spectroscopy would be able to a.) further confirm the Ca II K line matches the velocity of the H line cores b.) check for any Doppler variability in the Ca II K line and c.) check for additional, weaker metal lines at other wavelengths.

\subsection{IR Excess}
\label{irexcess}
We construct a spectral energy distribution (SED) using available catalogue photometry.  In the UV, we make use of the `GCAT' catalogue of unique \textit{GALEX} sources produced by M. Seibert and available at MAST\footnote{\url{https://archive.stsci.edu/prepds/gcat/}} as a High Level Science Product.  In the optical, we make use of the PanSTARRS \citep{kaiser2010} DR1 object catalogue \citep{flewelling2016} to get $g,r,i,z,y$ fluxes using the ``PSF'' magnitudes in the catalogue.  In the near-infrared, we make use of the VISTA Hemisphere Survey \citep[VHS,][]{mcmahon2013} to get $y, J, H, Ks$ fluxes.  We matched WD 2246-069 with sourceID 472553457122 in the DR4 release of the merged catalogue of sources, and used the available aperMag3 (2 arcsecond diameter) magnitudes.  In the mid-infrared, we make use of the AllWISE catalogue\footnote{\url{http://wise2.ipac.caltech.edu/docs/release/allwise/}}, which combines data from the cryogenic \citep{wright2010} and post-cryogenic \citep{mainzer2011} phases of the WISE mission.  Our source is detected in the \textit{W1} and \textit{W2} bands, where we use the available `w1pro' and `w2pro' magnitudes when analysing the target.  Only weak upper limits on the \textit{W3} and \textit{W4} fluxes are available, and are not used in our analysis.  Table \ref{wd2246_fluxes} summarizes the catalogue fluxes used in our analysis.

\begin{table}
\centering
\caption{Available catalogue fluxes used to analyse WD 2246-069's SED. See the description in Section \ref{irexcess} for references.\newline $^{\rm{a}}$95\% confidence limits determined by the WISE data reduction pipeline.}
\begin{tabular}{lllll}\label{wd2246_fluxes}
Band    	& System & $\lambda_{\rm{eff}}$ $(\rm{\mu m})$ & Mag. & Source \\\midrule
FUV			& AB	& $0.153$ & $18.348\pm0.037$ & GCAT \\
NUV			& AB	& $0.227$ & $17.492\pm0.013$ & GCAT \\
PS1\_g		& AB	& $0.487$ &	$16.880\pm0.008$ & PanSTARRS \\
PS1\_r		& AB	& $0.622$ &	$17.000\pm0.006$ & PanSTARRS \\
PS1\_i		& AB	& $0.755$ &	$17.250\pm	0.004$ & PanSTARRS \\
PS1\_z		& AB 	& $0.868$ &	$17.430\pm	0.010$ & PanSTARRS \\
PS1\_y		& AB 	& $0.963$ &	$17.532\pm	0.014$ & PanSTARRS \\
VHS\_y		& Vega & $1.020$ &	$16.988\pm	0.015$ & VHS \\
VHS\_J		& Vega	& $1.252$ &	$16.998\pm	0.020$ & VHS \\
VHS\_H		& Vega	& $1.645$ &	$17.010\pm	0.045$ & VHS \\
VHS\_Ks		& Vega & $2.147$ &	$17.061\pm	0.090$ & VHS \\
WISE\_1		& Vega	& $3.353$ & $16.750\pm	0.102$ & AllWISE \\
WISE\_2		& Vega	& $4.603$ &	$16.114\pm0.202$ & AllWISE \\
WISE\_3 	& Vega	& $11.56$ & $>12.473 ^{\rm{a}}$ & AllWISE \\
WISE\_4 	& Vega	& $22.09$ & $ > 8.860 ^{\rm{a}}$ & AllWISE \\
\bottomrule
\end{tabular}
\end{table}

\subsubsection{Possibility of Contamination}

\begin{figure}
\includegraphics[width=\linewidth]{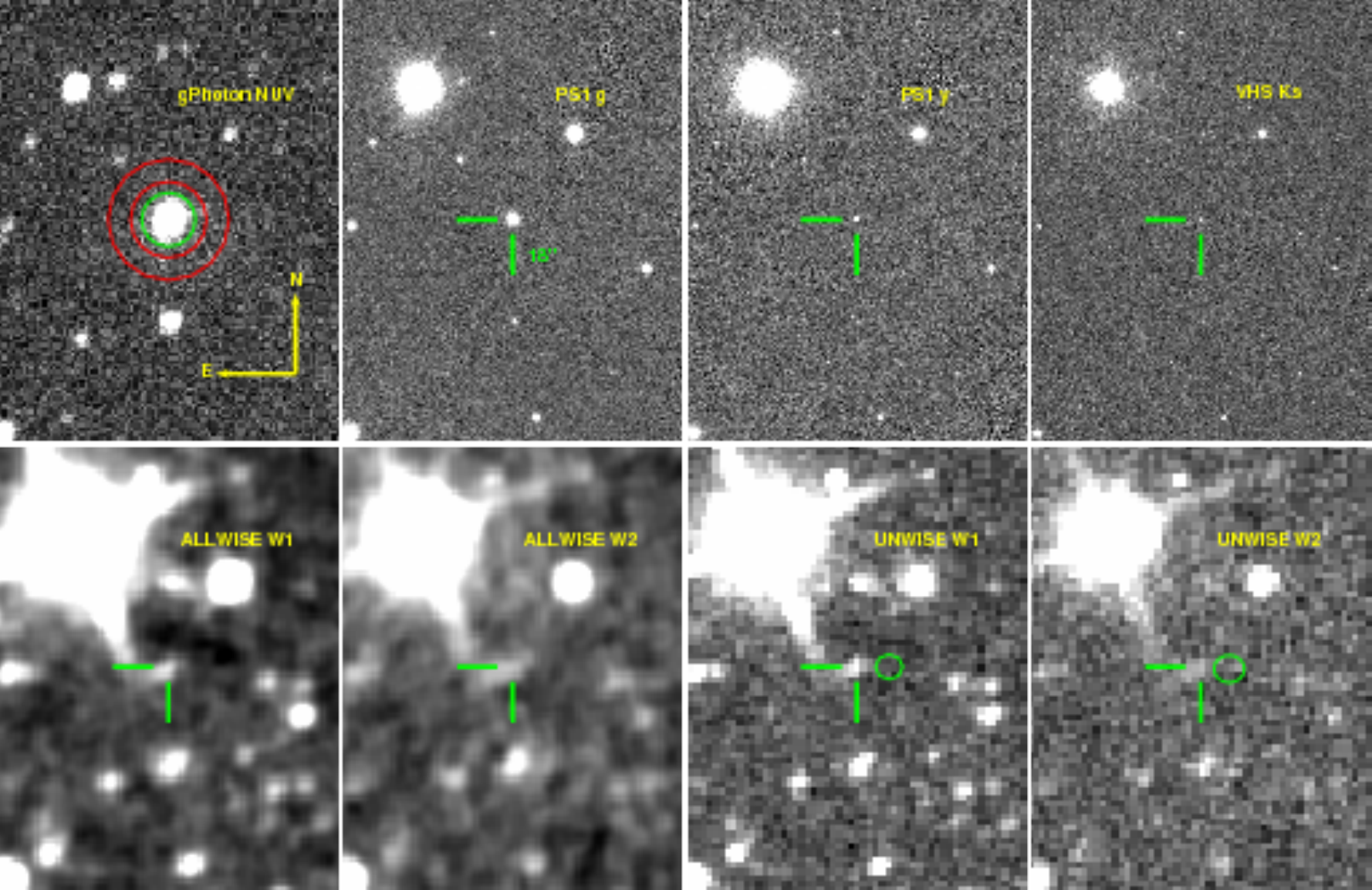}
\caption{Image atlas of WD 2246-069. Top row: \texttt{gPhoton} NUV counts, PanSTARRS $g$, PanSTARRS $y$, and VHS $Ks$. Bottom row: AllWISE \textit{W1}, AllWISE \textit{W2}, UNWISE \textit{W1}, and UNWISE \textit{W2}. The green and red circles in the \texttt{gPhoton} panel are the photometric aperture and background annuli used in gAperture.  The ellipses placed near the source in the UNWISE panels have radii equal to the FWHM of the WISE PSF in \textit{W1} and \textit{W2}. The AllWISE images are convolved using a model of these PSF's, a process that is not performed in the UNWISE images. The size of the green bars pointing to WD 2246-069 in each image are $18''$, for scale. A full resolution version of this image is available in the online manuscript.}
\label{2246imgatlas}
\end{figure}

A preliminary SED fit showed an excess in the two WISE bands at $\sim 3\sigma$.  This is significant, because only a few ZZ Ceti stars are known to have an IR excess (among them, G29-38 \citep{zuckerman1987, reach2005}, GD 133 \citep{jura2007}, WD 1150-153 \citep{kilic2007}, and PG 1541+651 \citep{kilic2012}).  Other possibilities are that it could be a background AGN, or an artefact in the WISE data.  We create image cutouts using the available data to check for any background companions that might be contaminating the photometry (Fig.\ \ref{2246imgatlas}).  The source appears extended in both the AllWISE and UNWISE \citep[unblurred WISE coadds,][]{lang2014,meisner2017} images, so it is possible the excess is due to an artefact, although we note the flag that marks extended sources is not set in the AllWISE catalogue, and the photometric quality flags for \textit{W1} and \textit{W2} are `A' and `B', respectively, indicating no major quality issues.  The WISE photometry was subject to the passive deblending routine, as tracked by the "nb" flag in the catalogue, to correct the photometry for blending from resolved sources further than 1.3 times the full-width-half-maximum.

The WISE (\textit{W1}-\textit{W2}) colour is redder than +0.3, thus WD 2246 lies beyond the cutoff used by \citet{hoard2013} to identify dusty WDs. While it is possible the WISE excess may be caused by a background object, this is unlikely for a few reasons.  The photo-centres of the catalogue positions were checked to ensure the WISE position is consistent with the optical sources.  The WISE coordinate is within 1.5 arcseconds of the SDSS and PanSTARRS positions, well within the WISE PSF.  The source is not extended in the Ks image from VHS, so if the WISE excess is caused by a background object it would have to be extremely faint in the near-IR.

We also check the possibility of a chance alignment with a background galaxy.  One of the more recent studies investigating the density of AGN in WISE data is \citet{assef2013}, who find a density for objects with $W2 < 17.11$ of $130 \pm 4 \; \rm{deg}^{-2}$, using the selection criteria $W1 - W2 \geq 0.8$. Encapsulating WD 2246-069 within a $15'' \times 15'' $ box, the probability of an AGN in this region is $\rm P \sim 0.2\%$. Since the AGN density determined by \citet{assef2013} extends to AGN with $W2 < 17.11$ and the measured $W2$ magnitude of WD 2246 is $\sim 16.1$, this probability is really an upper limit. A more stringent statistical limit could be placed using the non-detection of an extended source in the other bands, notably $Ks$. 

\begin{figure}
\includegraphics[width=\linewidth]{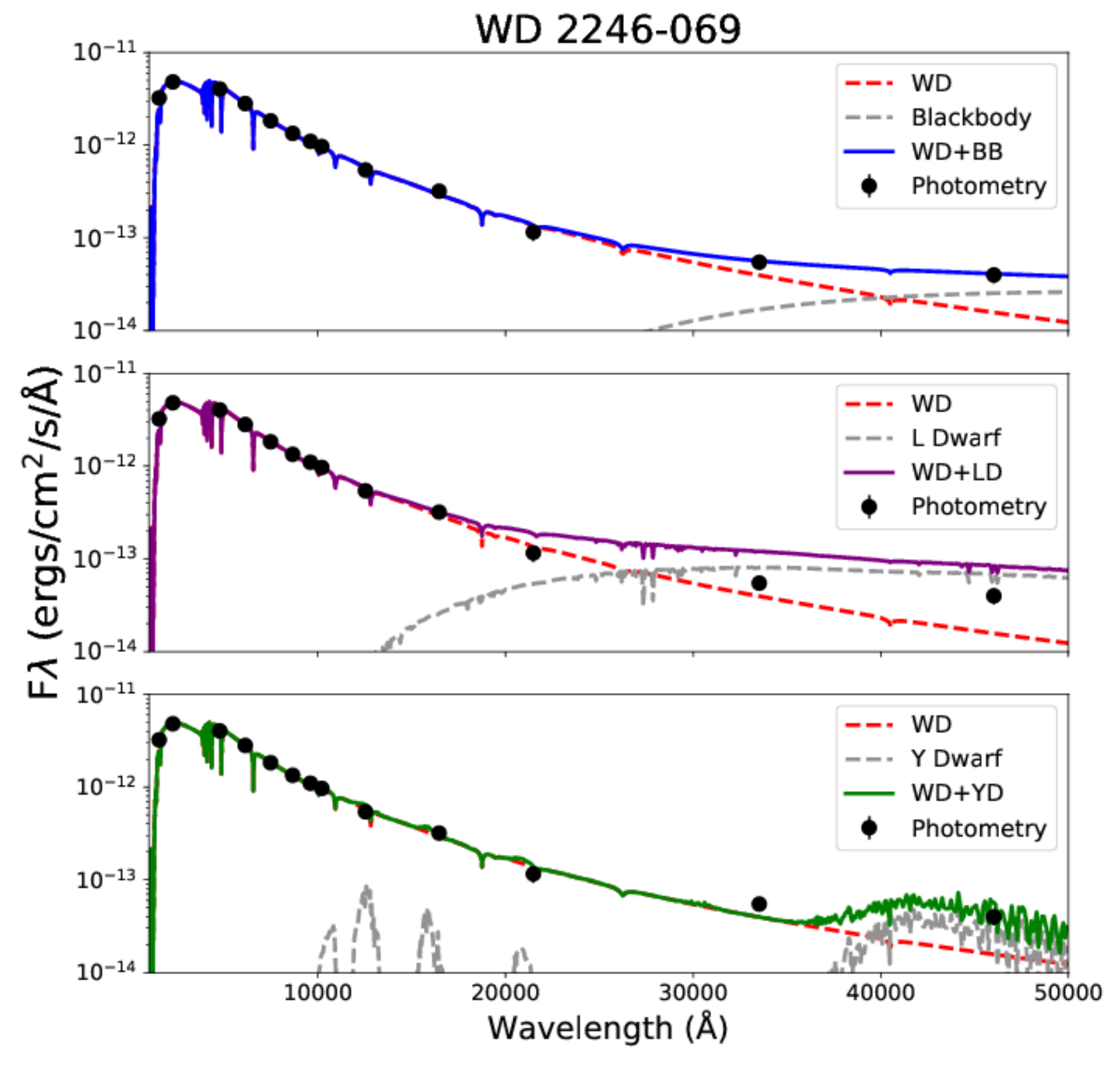}
\caption{\textit{Top}: Composite SED of a WD + blackbody (BB) component with $T_{\rm{eff}}\approx 670 \;\rm K$ and surface area $\sim 2 R_{\rm{jup}}$. This is the best fit to all of the photometry listed in Table \ref{wd2246_fluxes}. Note that the size of each photometry point is greater than or equal to the associated uncertainty in all the panels. \newline 
\textit{Middle}: Composite SED of a WD + L dwarf \citep[LD,][]{phoenixmodels} with $T_{\rm{LD}} = 1\, 000\, \rm K$ and an assumed $R_{\rm{LD}} = 1 R_{\rm{jup}}$ which is a conservative lower limit. The WD + LD SED contributes too much radiation in the WISE \textit{W1} and \textit{W2} bands to match the observed IR excess. \newline 
\textit{Bottom}: Composite SED of a WD + Y dwarf \citep[YD,][]{burrows03} companion with $ T_{\rm{YD}} = 686 \,\rm K$ and a model-computed radius of $R_{\rm{YD}} \sim 1.2 R_{\rm{jup}}$. The WD + YD SED matches the WISE \textit{W2} excess but does not contribute enough radiation in the \textit{W1} band to match the observed photometry.
}
\label{WD2246IRfit}
\end{figure}

\subsubsection{SED Fitting}

To investigate whether this excess could be caused by a faint companion, we fit the SED with synthetic models. To construct the composite SED, we make use of VOSA \citep{bayo2008} to convert the catalogue fluxes into units of $\rm{erg \; s^{-1} \; cm^{-2} \; \textrm{\AA}^{-1}}$.  The WD model is taken from an updated version of \citet{Koester2010}, where we select the model that is closest to the parameters derived from the asteroseismology.  We make use of BT-Dusty models \citep{phoenixmodels} and Y dwarf synthetic spectra \citep{burrows03} to add the spectra of low-mass, stellar secondary components and compare with the SED.  We find that an approximate distance of 50 pc results in a good match between the theoretical spectra and the observed fluxes, after scaling the models by $(r/d)^2$, where $r$ is the star's radius and $d$ is the distance to the system.  For the WD, we use a radius $r = 0.0107 \; R_{\odot}$, which is computed from the WD model to be consistent with a structure in equilibrium.  For the low-mass L dwarf companion, we assume $r = 1 \; R_{\rm{Jup}}$ as a canonical value. For the Y dwarf companion we use the computed model radius of $r \approx 1.2 \; R_{\rm{jup}}$.  We do not take into account extinction, but note that from the PanSTARRS 3D dust mapping \citep{green2015}, the estimated visual extinction at a distance of 100 pc is only $A_V \sim 0.01$, assuming a total-to-selective ratio $R_V = 3.1$.

Figure \ref{WD2246IRfit} shows the catalogue fluxes, the WD model, and three possible sources for the WISE excess: a blackbody component (top), a low-mass L dwarf companion (middle) and a Y dwarf companion (bottom).  We find a blackbody component with  $T_{\rm{eff}} \approx 670 \; \rm{K}$ and surface area $\sim 2 \; R_{\rm{jup}}$ matches the WISE excess well in both bands.  We conclude that it is unlikely that the WISE excess is caused by a bound L dwarf companion because the lowest mass model available \citep[$T_{\rm{eff}} = 1\, 000\; \rm K$, $\log g = 5.0$, solar abundances, ][]{phoenixmodels} contributes too much IR radiation in both WISE bands to match the observed fluxes.

Following the blackbody fit solution of $\approx 670 \, \rm K$, two Y dwarf models from \citep{burrows03} with similar $T_{\rm{eff}}$ are used to create synthetic composite spectra of a possible WD + Y dwarf system. The cooler Y dwarf spectra fit all observed photometry well with the exception of the WISE \textit{W1} band for which both models exhibit flux $2-3 \sigma$ below observed. The Y dwarf model presented in Fig. \ref{WD2246IRfit} corresponds to the $T_{\rm{eff}} = 686\; \rm K$ model but it is noted that the $T_{\rm{eff}} = 620 \; \rm K$ Y dwarf model also presents a similar match to the photometry. A valid scenario includes high mid-IR extinction in the ISM caused by dust grains, which is unaccounted for in \citet{green2015}, and might allow for a good fit with an L dwarf companion, but seems unlikely.

\section{Conclusion}
\label{conclusionsec}

We have conducted the first UV photometric WD survey for short time-scale variability using the \texttt{gPhoton} database and software.  We report the detection of pulsations in five WDs: four previously known from other studies and one new pulsator.  To our knowledge, these are the first image-based UV light curves for all these systems. There are two systems in this survey that stand out: WDs 1401-147 and 2246-069. WD 1401-147 is a unique case due to the near concurrent observations between the optical WET data from \citet{provencal2012} and our \textit{GALEX} UV data, which allows for direct UV-to-optical amplitude comparison.  The fit of the optical data was then used to predict the UV light curve and compared with the \texttt{gPhoton} data, with excellent agreement, confirming that the dominant spherical degree of the modes must be $\ell = 1$ or 2. The findings are discussed in \S\ref{wd1401sec}.

WD 2246-069 is a pulsating WD first identified in this survey from its \texttt{gPhoton} UV light curve. Optical data were taken at DSO and OPD to confirm its pulsating nature and perform asteroseismic analysis (Fig. \ref{wd2246_opd} \& \ref{wd2246_FTopd}, Table \ref{list-modes} \& \ref{best-fit}). Although WD 2246-069's $\log{g}$ and $T_{\rm{eff}}$ values are published in \citet{koester09}, placing it in the heart of the empirical instability strip (Fig. \ref{insta_strip}), no publications regarding its pulsation characteristics could be found. After confirming its pulsating nature, optical data from OPD was used to determine the star's independent modes and perform asteroseismology (Table \ref{list-modes} \& \ref{best-fit}).

Archival data show that this star also has a Ca II K absorption feature (Fig. \ref{2246spectra}), and an apparent mid-infrared excess in the WISE \textit{W1} and \textit{W2} bands (Fig. \ref{WD2246IRfit}). An intriguing explanation is the presence of a circumstellar debris disk/accretion source around the star.  Follow-up observations with better signal-to-noise spectra and/or additional data in the infrared will be able to confirm this scenario, which would add WD 2246-069 to the still very small list of ZZ Cetis that have both a disk and metal lines in their atmospheres.

As this star is one of just a few ZZ Ceti stars that possess both an IR excess matching that of a debris disk and metal lines in its optical spectrum, it provides unique insight into the interplay between a pulsating host star and its environment.  In particular, detailed modelling of the WD pulsations combined with infrared time-series photometry can constrain the geometry of the debris disk \citep{graham1990,patterson1991}. \citet{patterson1991} investigated the simultaneous photometric optical and IR variability of G29-38, the class prototype, finding some but not all of the WD's pulsations are also found in the mid-IR. This indicates that the debris disk responds to some of WD's pulsations, yet curiously not all of them. One hypothesized explanation for the selective nature of the IR pulsations would require $m=0$ for the IR-detected modes \citep{graham1990}, however, \citet{patterson1991} find that a very unusual geometric orientation is needed to confirm this. 

Besides IR photometric studies, time-series spectroscopy of pulsating WD with metal lines can use the variations in the EW of these metal lines to determine the accretion geometry of the system, which has only been applied to a single system by \citet{montgomery08} and \citet{thompson10}, who interestingly found conflicting results. Further work expanding this sample is necessary for completing our understanding of WD accretion scenarios and possible progenitors.

Additionally, DAVs and pulsating WDs in general are useful in probing the interior structure of WD stars. Since the conditions of WD interiors cannot be recreated in the laboratory at this time, these stars allow the study of degenerate stellar material and its properties, especially with relation to energy transport mechanisms. A comprehensive overview of the physics involved is given by \citet{althaus10} with research examples including \citet{althaus2010a} and \citet{romero17}.  

We have presented a new way to discover variable WDs using the gPhoton database and software. While this project studied a subsample of the Mccook-Sion catalog, similar searches for pulsating white dwarfs using other WD catalogues as a starting point, such as SDSS, can yield further insight.

\section*{Acknowledgements}

The authors would like to thank Richard O. Gray for his comments on the manuscript, and Eilat Glikman \& Dave Sanders for their suggestions regarding WISE AGN densities. 

Some of the data presented in this paper were obtained from the Mikulski Archive for Space Telescopes (MAST). STScI is operated by the Association of Universities for Research in Astronomy, Inc., under NASA contract NAS5-26555. Support for MAST for non-HST data is provided by the NASA Office of Space Science via grant NNX09AF08G and by other grants and contracts.

This research has made use of the NASA/ IPAC Infrared Science Archive, which is operated by the Jet Propulsion Laboratory, California Institute of Technology, under contract with the National Aeronautics and Space Administration.  Based on observations obtained as part of the VISTA Hemisphere Survey, ESO Program, 179.A-2010 (PI: McMahon).  This publication makes use of VOSA, developed under the Spanish Virtual Observatory project supported from the Spanish MICINN through grant AyA2011-24052.  This research has made use of the SIMBAD database \citep{wenger2000}, operated at CDS, Strasbourg, France.   This research has made use of the VizieR catalogue access tool, CDS, Strasbourg, France. The original description of the VizieR service was published in \citet{ochsenbein2000}.  This publication makes use of data products from the Wide-field Infrared Survey Explorer, which is a joint project of the University of California, Los Angeles, and the Jet Propulsion Laboratory/California Institute of Technology, funded by the National Aeronautics and Space Administration.

The Pan-STARRS1 Surveys (PS1) and the PS1 public science archive have been made possible through contributions by the Institute for Astronomy, the University of Hawaii, the Pan-STARRS Project Office, the Max-Planck Society and its participating institutes, the Max Planck Institute for Astronomy, Heidelberg and the Max Planck Institute for Extraterrestrial Physics, Garching, The Johns Hopkins University, Durham University, the University of Edinburgh, the Queen's University Belfast, the Harvard-Smithsonian centre for Astrophysics, the Las Cumbres Observatory Global Telescope Network Incorporated, the National Central University of Taiwan, the Space Telescope Science Institute, the National Aeronautics and Space Administration under Grant No. NNX08AR22G issued through the Planetary Science Division of the NASA Science Mission Directorate, the National Science Foundation Grant No. AST-1238877, the University of Maryland, Eotvos Lorand University (ELTE), the Los Alamos National Laboratory, and the Gordon and Betty Moore Foundation.

K.J.B. and M.H.M. acknowledge support from NSF grant AST-1312983. I.P. and A.D.R. acknowledge support from CNPq-Brazil.

\nocite{*}
\bibliographystyle{mnras}
\bibliography{references}



\label{lastpage}

\end{document}